\documentclass[pre,a4paper,onecolumn,noshowpacs,showkeys,nofootinbib,floatfix]{revtex4}
\usepackage{graphics}
\usepackage{graphicx}
\usepackage{epsfig}
\usepackage{amssymb}
\begin{document}
\title{Effective multifractal features and $ \ell $-variability diagrams of
high-frequency price fluctuations time series}
\thanks{The main contents of this manuscript is part of a CBPF document by
JdS publicly presented at this institute on the $12^{th}$ February 2007.}
\author{Jeferson de Souza}
\email{jdesouza@ufpr.br}
\affiliation{Laborat\'{o}rio de An\'{a}lise de Bacias e Petrof\'{\i}sica, Departamento de Geologia, \\ 
Universidade Federal do Paran\'{a}, Centro Polit\'{e}cnico - Jardim das Am\'{e}ricas, \\ 
Caixa Postal 19001, 81531-990 Curitiba-PR, Brazil 
\\ and \\
Centro Brasileiro de Pesquisas F\'\i sicas, \\ 
Rua Dr. Xavier Sigaud 150, 22290-180 Rio de Janeiro-RJ, Brazil}

\author{S\'{\i}lvio M. Duarte~Queir\'{o}s}
\altaffiliation[Present address: ]{Unilever R\&D Port Sunlight, Quarry Road East, Wirral CH63 3JW, United Kingdom}
\email{sdqueiro@cbpf.br;sdqueiro@googlemail.com}
\thanks{Corresponding author.}
\affiliation{Centro Brasileiro de Pesquisas F\'\i sicas, \\ 
Rua Dr. Xavier
Sigaud 150, 22290-180 Rio de Janeiro-RJ, Brazil}

\begin{abstract}
In this manuscript we present a comprehensive study on the multifractal
properties of high-frequency price fluctuations and instantaneous volatility
of the equities that compose Dow Jones Industrial Average. The analysis
consists about quantification of dependence and non-Gaussianity on the
multifractal character of financial quantities. Our results point out an 
equivalent influence of dependence and non-Gaussianity on the multifractality of time series. 
Moreover, we analyse $\ell $-diagrams of price fluctuations. 
In the latter case, we show that the fractal dimension of these maps is basically independent of the 
lag between price fluctuations that we assume.
\end{abstract}
\maketitle

\section{Introduction\label{intro}}

Scale invariance and fractality, \textit{i.e.}, the absence of a
characteristic scale can be found in a widespread of natural and human
creation phenomena \cite{mandelbrot}. Mathematically, scale invariance of a
certain function, $f$, of an observale $\mathcal{O}$ is written as,%
\begin{equation}
f\left( \lambda \mathcal{O}\right) =\lambda ^{\alpha }f\left( \mathcal{O}%
\right) ,  \label{scaleinv}
\end{equation}%
and it has consensually been considered as a signature of complexity \cite%
{complex}. Specifically, power-law behaviour respecting Eq. (\ref{scaleinv})
has empirically been verified in the probability density and autocorrelation
functions of several time series such as: fluctuation in heart rate beating 
\cite{peng-heart}, gait \cite{gait}, variation in the magnetic field of the
solar wind in the heliosheath \cite{nasa}, or relative stock price
fluctuations \cite{bouchaud}\cite{mantegna} among many others. Concerning
time series and fractality \footnote{%
Time series can only be considered scale invariante in a self-affine context.%
}, if many of them seem to be \textit{monofractal} \cite{feder}, \textit{i.e.%
},\ they are characterised by a single scale exponent, just as in Eq.(\ref%
{scaleinv}), other time series, namely those we have referred to here above,
have shown a spectrum of locally dependent $\alpha $ exponents.
Analytically, this is noted as%
\begin{equation}
f\left( \left\{ \lambda \mathcal{O}\right\} ^{\upsilon }\right) =\lambda
^{\alpha \left( \upsilon \right) }f\left( \mathcal{O}^{\upsilon }\right) .
\label{scaleinvmulti}
\end{equation}%
The previous Eq. (\ref{scaleinvmulti}) is also valid for multiscaling and 
\textit{multifractality} as well, which has consistently been associated
with the main statistical features of time series obtained from complex
systems. Consequently, this close relation has been prominent in either the
development of dynamical models or validation of previous approaches. In the
former case, pioneering works by \textsc{B. Mandelbrot} have opened the door
to a new treatment of financial markets dynamics \cite{mandelbrot-scaling}.

In sequel of this manuscript we perform an extensive analysis of the
statistical features of high-frequency price fluctuations, $r_{t}$, of the $%
30$ equities that compose the Dow Jones Industrial Average (DJIA). Previous
studies on daily price fluctuations have shown the existence of a
multifractal behaviour \cite{djia-daily}. Hence, with this high-frequency
analysis, it is our aim to study the multiscaling of price fluctuations at a
level that is closer to the transaction dynamics as it has been made for
other financial observables. Our study is driven on the evaluation of the
multifractal spectra of both of time series and $\left( \ell =1\right) $%
-diagrams, $\left( r_{t},r_{t+1}\right) $, describing their main factors of
multifractality. In addition, we enquire into price fluctuations absolute
values, $\left\vert r_{t}\right\vert $, also called as \emph{instantaneous\
volatility}, $v_{t}\equiv \left\vert r_{t}\right\vert $, multifractal
behaviour and analyse its weight on the multiscaling characteristics of
price fluctuations. Our manuscript is organised as follows: in Sec. \ref%
{data-method} we describe the data used and the methodology applied in order
to obtained multifractal spectra for time series and $\ell $-diagrams. Along
Sec. \ref{res-time-series} we present our results of the analysis of price
fluctuations and volatility multifractal spectra. This comprises the
quantification of the key elements of multiscaling for both quantities. In
addition, we verify the plausibility of a superstatistical approach (which
is able to provide a nice answer within the context of price fluctuations
probability density function) in a multifractal characterisation of price
fluctuations. In Sec. \ref{l-diagrams} we present the results of the study
of the fractal dimension of price fluctuations $\ell $-diagrams. To
finalise, some remarks, conclusions, and perspectives for future work are
set forth in Sec. \ref{remarks}.

\section{Data and methods\label{data-method}}

\subsection{Data}

Our data is composed by $1$ minute time series of the prices, $S_{i}\left(
t\right) $ ($i$ stands for the company), of the $30$ companies that composed
the Dow Jones Industrial Average from the $1^{st}$ of July until the $%
31^{st} $ December of $2004$ in a total of around $50\,000$ data points for
each equity. For each $i$ equity we have computed $1$ minute ($\log $-)price
fluctuations as,%
\begin{equation}
\tilde{r}_{i}^{\prime }\left( t\right) =\ln S_{i}\left( t\right) -\ln
S_{i}\left( t-1\right)  \label{return1}
\end{equation}%
It is well known that trading activity exhibits a intraday pattern \cite%
{admati}. In other words, markets tend to be highly active (hence volatile)
in the first $30$ minutes of a business day, mainly to take advantage from
news and events between the closure of the previous market session and the
next following opening. After a decrease of activity along the day, markets
present an activity set-up in the final part of trading sessions, basically
due to the action of liquid traders. This \textit{U-shape} enhances spurious
features namely in correlations. To remove it, we have performed according
to the following standard procedure:

\begin{itemize}
\item After we have computed $1$ minute price fluctuations, as in Eq. (\ref%
{return1}), we have determined the average volatilities, $\Lambda $,
associated with equity $i$ and intra-day time $t^{\prime }$ (which has an
upper bound of $340$ minutes for companies traded at $NYSE$ and $420$
minutes for companies traded at $NASDAQ$),%
\begin{equation}
\Lambda _{i}\left( t^{\prime }\right) =\frac{\sum_{j=1}^{N}\left\vert \tilde{%
r}_{i}^{\prime }\left( t^{\prime },j\right) \right\vert }{N},
\label{intra-volatility}
\end{equation}%
where $N$ represents the number of days for which the market was trading at $%
t^{\prime }$ intra-day time;

\item We have then used the average volatilities to normalise price
fluctuations, eliminating the intraday pattern,%
\begin{equation}
\tilde{r}_{i}\left( t\right) \rightarrow \frac{\tilde{r}_{i}\left( t^{\prime
}\right) }{\Lambda _{i}\left( t^{\prime }\right) },  \label{return-semintra}
\end{equation}%
where we have dropped the prime of $t$ in the left-hand side, because the
time series has lost its intra-day profile;

\item To complete, we have removed the average and normalised $\left\{ 
\tilde{r}_{i}\left( t\right) \right\} $ by its standard deviation,%
\begin{equation}
r_{i}\left( t\right) =\frac{\tilde{r}_{i}\left( t\right) -\left\langle 
\tilde{r}_{i}\right\rangle }{\sqrt{\left\langle \left( \tilde{r}_{i}\right)
^{2}\right\rangle -\left\langle \tilde{r}_{i}\right\rangle ^{2}}},
\end{equation}%
($\left\langle \ldots \right\rangle $ represents average over all the
elements of time series $i$), to define our studied price fluctuations, $%
r_{i}\left( t\right) $.
\end{itemize}

We next present our methods to evaluate multifractal spectra for time
series, multifractal detrended fluctuation analysis (MF-DFA), and $\ell $%
-diagrams.

\subsection{MF-DFA}

The MF-DFA \cite{mf-dfa}\ is one of the most applied methods to determine
the multifractal properties of time series in several fields \cite%
{mf-dfa-applications}\cite{canberra}. We have chosen to apply MF-DFA in \
lieu of Wavelet Transform Modulus Maxima (WTMM) \cite{wavelet} taking into
account a recent comparative study where it has been shown that in the
majority of situations MF-DFA presents reliable results \cite{polacos},
i.e., it does not introduce specious multifractality, at least in the
amounts that have been computed from WTMM. The MF-DFA method goes as follows:

Consider the time series $\left\{ x_{i}\left( t\right) \right\} $ ($x_{i}$
represents both of price fluctuations and instantaneous volatilities of
company $i$) composed by $N$ ($N\gg 1$),

\begin{itemize}
\item Determine the profile $Y_{i}\left( t\right) $ that corresponds to the
deviation of signal elements from the mean%
\begin{equation}
Y_{i}\left( t\right) =\sum_{l=1}^{t}\left[ x_{i}\left( t\right)
-\left\langle x\right\rangle \right] ,\qquad \left( 1\leq t\leq N\right) ,
\end{equation}%
and thereafter,%
\begin{equation}
\tilde{Y}_{i}\left( t\right) =\sum_{l=1}^{t}\left[ Y_{i}\left( t\right)
-\left\langle Y_{i}\right\rangle \right] ;
\end{equation}

\item Divide profile $\tilde{Y}_{i}\left( t\right) $ into $N_{s}\equiv
int\left( \frac{N}{s}\right) $ non-overlapping intervals of equal size $s$;

\item Compute local tendency by a least-square adjustment, and thereupon
variance\footnote{%
In the rest of this section we omit company index $i$ to turn out notation
lighter.},%
\begin{equation}
\tilde{F}^{2}\left( \nu ,s\right) =\frac{1}{s}\sum_{l=1}^{s}\left\{ \tilde{Y}%
\left[ \left( \nu -1\right) \,s+l\right] -y_{\nu }\left( l\right) \right\}
^{2},
\end{equation}%
for each segment $\nu $, $\nu =1,\ldots ,N_{s}$, where $y_{\nu }\left(
i\right) $ represents a $m^{th}$-order polynomial. The order of the
polynomial is relevant on the results one might obtain. For the series we
have analysed we have used polynomials of order $5$ from which on we could
not appraise changes of the values of multifractal spectra.

\item Figure out the average $F_{z}\left( s\right) $ over all segments to
obtain the fluctuation function of order $z$,%
\begin{equation}
F_{z}\left( s\right) \equiv \left\{ \frac{1}{N_{s}}\sum_{\nu =1}^{N_{s}}%
\left[ \tilde{F}^{2}\left( \nu ,s\right) \right] ^{z/2}\right\}
^{1/z},\qquad \forall _{z\neq 0},
\end{equation}%
and%
\begin{equation}
F_{z}\left( s\right) \equiv \exp \left\{ \frac{1}{2N_{s}}\sum_{\nu
=1}^{N_{s}}\ln \left[ \tilde{F}^{2}\left( \nu ,s\right) \right] \right\}
,\qquad z=0.
\end{equation}

\item Assess the scaling behaviour of $F_{z}\left( s\right) $ considering $%
\log -\log $ scale representation of $F_{z}\left( s\right) $ \textit{vs}. $s$
for each value of $z$. In case the series $\left\{ x\left( t\right) \right\} 
$ shows multiscaling features then,%
\begin{equation}
\frac{F_{z}\left( s\right) }{s}\sim s^{h\left( z\right) }.  \label{mf-dfa1}
\end{equation}
\end{itemize}

Small fluctuations are generally characterised by large scale values of
exponent $h\left( z\right) $ (and $z<0$), whereas large fluctuations are
typified by small values of $h\left( z\right) $ (and $z>0$).

To bridge this procedure with the standard formalism, we can verify that $%
\left[ F_{z}\left( s\right) \right] ^{z}$ can be interpreted as the
partition function, \ $Z_{z}\left( s\right) $ \cite{feder}, which is known
to scale with the size of the interval as,%
\begin{equation}
Z_{z}\left( s\right) \sim s^{\tau \left( z\right) }.  \label{mf-dfa2}
\end{equation}%
Hence, according to Eq. (\ref{mf-dfa1})\ and Eq. (\ref{mf-dfa2}) we have,%
\begin{equation}
\tau \left( z\right) =z\,h\left( z\right) -1.
\end{equation}%
Using Legendre transform,%
\begin{equation}
f\left( \alpha \right) =z\,\alpha -\tau \left( z\right) ,  \label{mf-dfa-tau}
\end{equation}%
we can relate exponent $\tau \left( z\right) $ with H\"{o}lder exponent, $%
\alpha $, 
\begin{equation}
\alpha =h\left( z\right) +z\frac{dh\left( z\right) }{dz},
\label{mf-dfa-alfa}
\end{equation}%
and 
\begin{equation}
f\left( \alpha \right) =z\,\left[ \alpha -h\left( q\right) \right] +1.
\label{mf-dfa-falfa}
\end{equation}%
For $z=2$, $h\left( 2\right) \equiv H$, that corresponds to the Hurst
exponent \cite{hurst} customarily determined by methods like the original $%
R/S$ ratio or DFA \cite{dfa} (for nonstationary signals) from which MF-DFA
derives. For $z=0$, $f\left( \alpha \right) $ obtained from Eq. (\ref%
{mf-dfa-tau}), Eq. (\ref{mf-dfa-alfa}), and Eq. (\ref{mf-dfa-falfa})
corresponds to the support dimension.

In the case of a monofractal, $h\left( z\right) $ is independent from $z$,
since there is homogeneity in the scaling behaviour. Specifically, there
exist only different values of $h\left( z\right) $, for each $z$, if large
and small fluctuations scale in different ways.

\subsection{Box counting algorithm \label{hou-method}}

Box-counting methods have been extensively applied to determine scaling (fractal) properties of measures. 
As a matter of fact, it is the standard procedure to verify the factal nature of measures~\cite{feder}. 
To determine multiscaling properties in the $\ell $-diagrams we have
used a recently presented optimised implementation~\cite{petrobras} of the procedure introduced in Ref. \cite{hou}. 
This method is an improvement of the algorithm proposed by Liebovitch and Toth \cite{lieb},
and it is based on the fact that coordinates of a fractal, suitably shifted
and rescaled, and written on the binary numerical base can be combined to
form bit strings with $k\,D_{E}$ bits whose first $m\,D_{E}$ bits from left
to right determine uniquely the position of the coordinates in $D_{E}$%
-dimensional space. Here, $m=1,2,\cdots ,k$ and $k$ is a positive integer.
Thus, the $N\,D_{E}$ coordinates are mapped in $N$ bit strings and with $%
k\,D_{E}$ bits, where $k$ is the maximal number of bits used to represent
each coordinate on the binary base. After masking $m\,D_{E}$ bits from right
to left, strings that have the same position code belong to the same box in
resolution $m$. Then, by scanning the $N$ bit strings, the number of changes
are stored. The number of changes represents the number of boxes needed to
cover the fractal set in the scale $m=\ln _{2}\,s$. If the set presents a
fractal measure, one has%
\begin{equation}
N\sim s^{-D_{f}},  \label{boxcount}
\end{equation}%
where $D_{f}$ is the fractal dimension of the set.

\subsection{Quantification of multiscaling components}

As we have stated in Sec.~\ref{intro} there has been established a close
relation between multiscaling and both of correlations and probability
density functions of price fluctuations time series. In another perspective,
we can ascribe to memory and non-Gaussianity of probability density
functions the emergence of multifractal characteristics in price
fluctuations~\cite{mf-dfa}. This can be made if we first consider such
contributions as independent. Upon this assumption, we can quantify their
relative weights. In other words, if we aim to size up the weight of
non-Gaussianity we must destroy memory in the signal. And from it, by using
the independence conjecture, we determine memory influence. On one hand,
memory is basically destroyed if we shuffle time series elements. Doing
that, we reorder the values of our original time series, but we keep the
stationary probability density function unalterable. On the other hand, we
can destroy non-Gaussianity by implementing the procedure which we call as 
\textit{phase randomisation}:

\begin{itemize}
\item Determine the Fourier transform of the signal $\left\{ x\left(
t\right) \right\} $, 
\begin{equation}
\xi _{f}\equiv \mathcal{F}\left[ x\left( t\right) \right] ;
\end{equation}

\item Dissociate amplitude from the phase of the transformed signal,%
\begin{equation}
\xi _{f}=\left\vert \xi _{f}\right\vert \exp \left[ i\,\arctan \frac{\mathrm{%
Im}\left( \xi _{f}\right) }{\mathrm{Re}\left( \xi _{f}\right) }\right] ;
\end{equation}%
where \textrm{Im}(\textrm{Re}) stands for \textit{imaginary}(\textit{real})
part of some complex number $\xi _{f}$.

\item Expunge the phase of the transformed signal and introduce new random
phases, $\theta _{rnd}$, uniformly distributed, for half of the elements,
and assign for the other half of the series a phase $-\theta _{rnd}$. In
this way we define a phase randomised signal, $\xi _{f}^{rnd}$,%
\begin{equation}
\xi _{f}^{rnd}=\left\vert \xi _{f}\right\vert \exp \left[ i\,\theta _{rnd}%
\right] ;
\end{equation}

\item Apply the inverse Fourier transform on the phase randomised signal, 
\begin{equation}
x^{rnd}\left( t\right) \equiv \mathcal{F}^{-1}\left[ \xi _{f}^{rnd}\right] .
\end{equation}
\end{itemize}

For both shuffled and phase randomised time series obtained from the
original signal, we can also carry out a MF-DFA analysis. For each case, Eq.
(\ref{mf-dfa1}) can be verified where we use exponents $h_{shf}\left(
z\right) $ for the shuffled time series, and $h_{rnd}\left( z\right) $ for
the phase randomised case. Assuming independency between multifractal
factors, we have measured the contribution of correlations, $h_{cor}\left(
z\right) $, by,%
\begin{equation}
h_{cor}\left( z\right) \equiv h\left( z\right) -h_{shf}\left( z\right) .
\end{equation}%
If only these two factors introduce multiscaling on the signal then, when we
perform the phase randomisation process on a shuffled signal, we should
obtain a Gaussian and uncorrelated signal, \textit{i.e.}, $h_{shf-rnd}\left(
z\right) =\frac{1}{2}$ for all $z$. Theoretically, we can evaluate the
contribution of non-Gaussianity, $h_{PDF}^{\prime }\left( z\right) $, from
phase randomised time series as well,%
\begin{equation}
h_{PDF}^{\prime }\left( z\right) =h_{shf}\left( z\right) -h_{shf-rnd}\left(
z\right) .
\end{equation}

However, the probability density function of a finite time series is
influenced by its size, particularly for small time series \cite{kruger}. In
this sense, comparing results obtained from times series with different
probability density functions, such is the case of $\left\{ x^{shf}\left(
t\right) \right\} $ and $\left\{ x^{shf-rnd}\left( t\right) \right\} $,
introduces error factors that we are not able to quantify. Regarding this
factor, we have opted to define an effective contribution of
non-Gaussianity, $h_{PDF}\left( z\right) $, 
\begin{equation}
h_{PDF}\left( z\right) =h_{shf}\left( z\right) .
\end{equation}%
In a previous article by us~\cite{bariloche}, in which we analyse the
multifractal features of traded volume for the same equities, we have
computed a contribution that we have called as non-linear effects. These
effects are actually governed by finite size effects that play a significant
role on the multifractal character of a time series when large values of $%
\left\vert z\right\vert $ are taken into account. Moreover, the finiteness
of a time series might introduce fake multiscaling features. This fact
emphasises the sensitiveness of multifractal measurements which are many
time inflated by artefacts. In order to avoid, or at least minimise those
spurious features, a careful choice of the range of $s$ and $z$ values must
be made. In our analysis, we have chosen $s$ between $8$ and $11585$, and $z$
between $-3$ and $5$. Within this range of values we were able to obtain
numerical curves which concur to the theoretical scaling curve of
independent and Gaussian time series.

Multifractality can be effectually quantified through the difference between
scale exponents of $z_{\min }$\ and $z_{\max }$,%
\begin{equation}
\Delta h\equiv h\left( z_{\min }\right) -h\left( z_{\max }\right) .
\label{deltah}
\end{equation}%
For a monofractal, $\Delta h=0$, because of the linear dependence of $\tau $
with $z$. Equation~(\ref{deltah}) can be used for the original time series, $%
\Delta h$, and for the shuffled time series, $\Delta h_{shf}$. From these
values, we finally compute the weight of non-Gaussianity, $\Delta
h_{shf}/\Delta h$, and of correlations $1-\Delta h_{shf}/\Delta h$.

\section{Results for time series\label{res-time-series}}

\subsection{Multifractality for price fluctuations time series}

\begin{figure}[th]
\centering 
\includegraphics[width=0.75\columnwidth,angle=0]{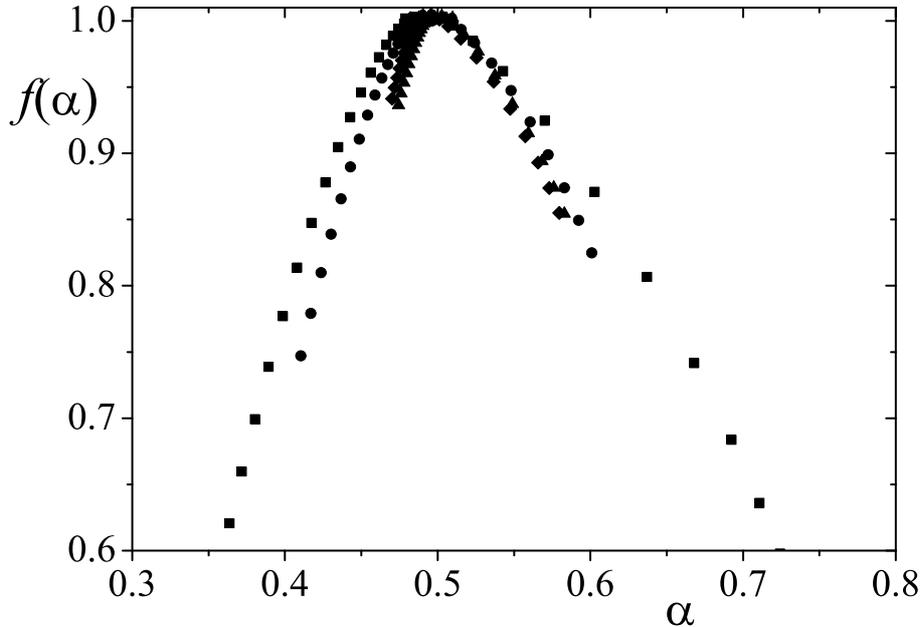}
\caption{ Multifractal spectra $f$ \textit{vs.} $\protect\alpha $ of the
price fluctuations ($\blacksquare $), shuffled time series ($\bullet $),
phase randomised ($\blacklozenge $), and shuffled plus phase randomised ($%
\blacktriangle $) time series of DJIA equities. As it is depicted, when
elements that introduce multiscaling are removed the multifractal spectrum
becomes narrower. }
\label{fig-ret-falfa}
\end{figure}

In Fig.~\ref{fig-ret-falfa} we present our results for the multifractal
spectrum of price fluctuations, shuffled, phase randomised, and shuffled
plus phase randomised time series. The values have been obtained by
performing an average over the $30$ companies of moments $\tau $. Despite of
the fact that it has been verified the influence of equities liquidity on
the multifractal properties of financial time series, all companies of our
data set have presented liquidity values within the same order of magnitude,
turning out our average over the companies perfectly plausible. As it can be
seen, the price fluctuations time series present a wide multifractal
spectrum with $\alpha _{\min }=0.364$ and $\alpha _{\max }=0.724$.
Furthermore, we verify a strong asymmetry between the part of the spectrum
that goes from $\alpha _{\min }$ up to $\alpha \left( z=0\right) $ and the
remaining part of the spectrum. The asymmetry in Fig. \ref{fig-ret-falfa} is
contrary to the $f\left( \alpha \right) $ curve that has been measured in
fully developed turbulent flows \cite{meneveau} often considered a price
fluctuation analogue. Concerning the other time series, we observe that the
multifractal spectrum of the shuffled time series is less wider than the
spectrum for the original time series. In addition, the shuffled signals
have larger spectrum than the randomised and shuffled plus phase randomised
signals. Analysing scaling exponent $h\left( 2\right) $, that is the common
Hurst exponent, $H$, we have obtained a value around $\frac{1}{2}$
concomitant with a white noise sequence, and in accordance with the
Efficient Market Hypothesis (EMH) \cite{fama}. Furthermore, we have $\max
\left( f\left( \alpha \right) \right) =1$, \textit{i.e.}, price fluctuations
time series are \textit{fat-fractals} as it occurs for a large variety of
other signals and non-linear phenomena \cite{mandelbrot}\cite{farmer-fat}.
For the $h$ difference defined in Eq. (\ref{deltah}) we have obtained $%
\Delta h=0.15$ and $\Delta h_{shf}=0.08$. These values yield a weight of $%
54\%$ for non-Gaussianity and $46\%$ for correlations in the multifractal
properties of our time series. In spite of this result appears to be at odds
with the $H=\frac{1}{2}$, we must call attention to the fact that there is a
more delicate relation for random variables, \emph{the statistical dependence%
} \cite{feller}, which cannot be described by the Hurst exponent. The
statistical dependence of financial observables \cite{bouchaud}\cite%
{serletis}\cite{smdq-qf} has been verified by means of mutual information
measures \cite{dependence}. We attribute to this statistical feature the
multiscaling of price fluctuations we have perceived. This assignment is
also supported by the structure of $\left( \ell =1\right) $-diagrams that we
analyse in Sec.~\ref{l-diagrams}. 

\begin{figure}[th]
\centering 
\includegraphics[width=0.4\columnwidth,angle=0]{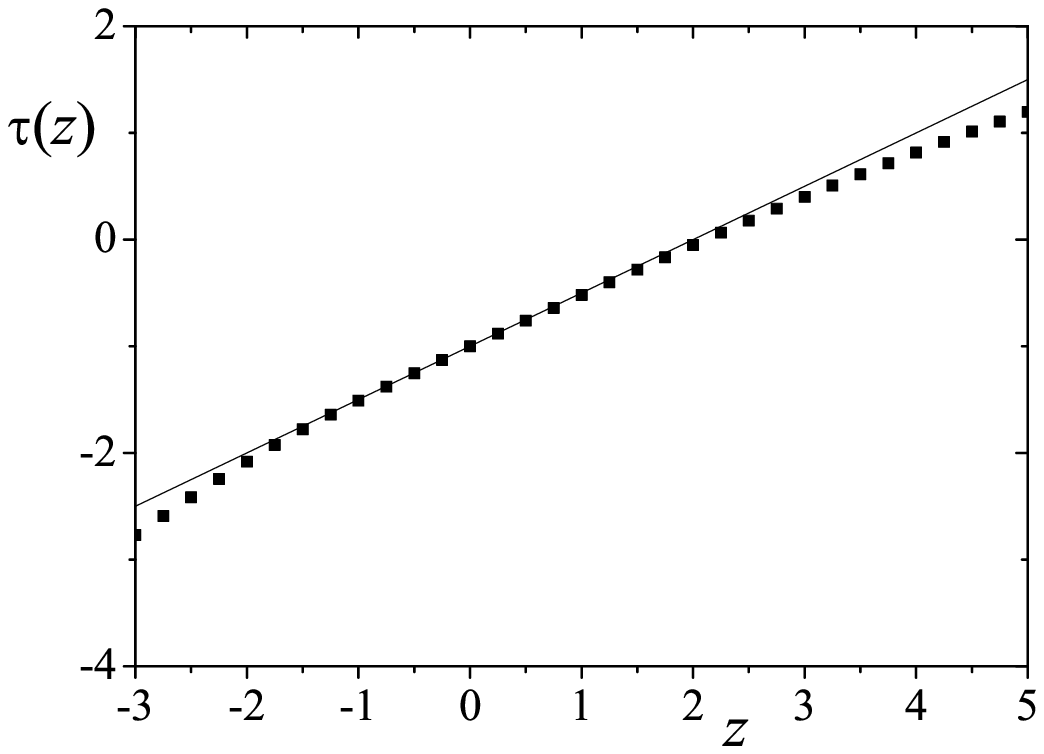} %
\includegraphics[width=0.4\columnwidth,angle=0]{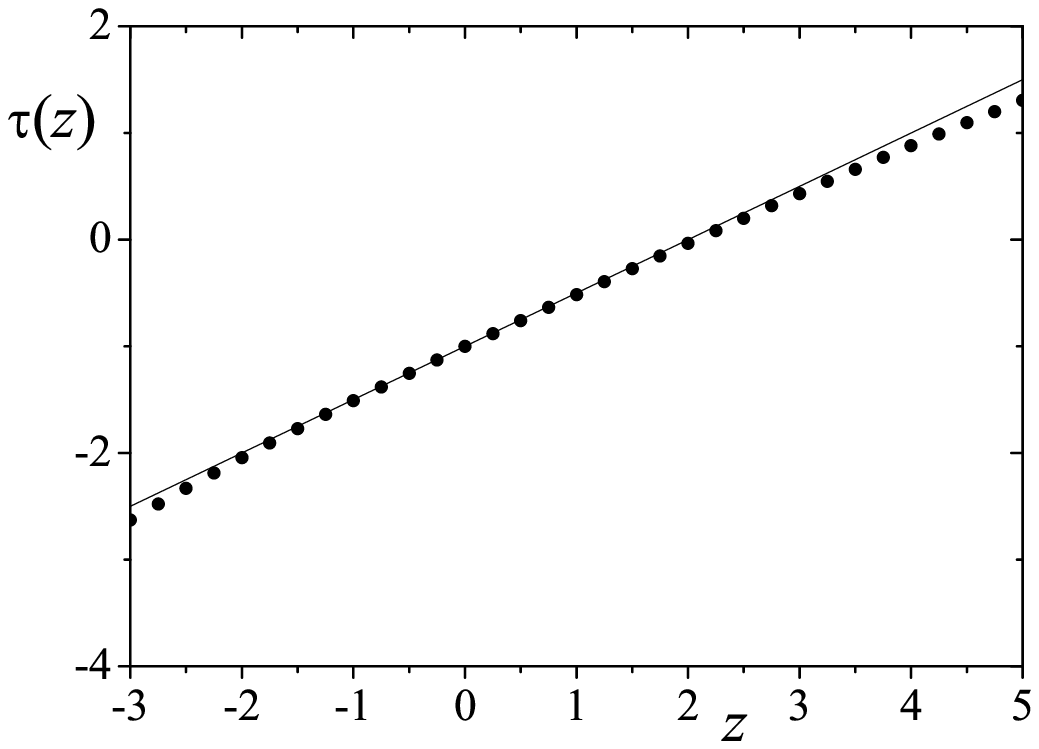} %
\includegraphics[width=0.4\columnwidth,angle=0]{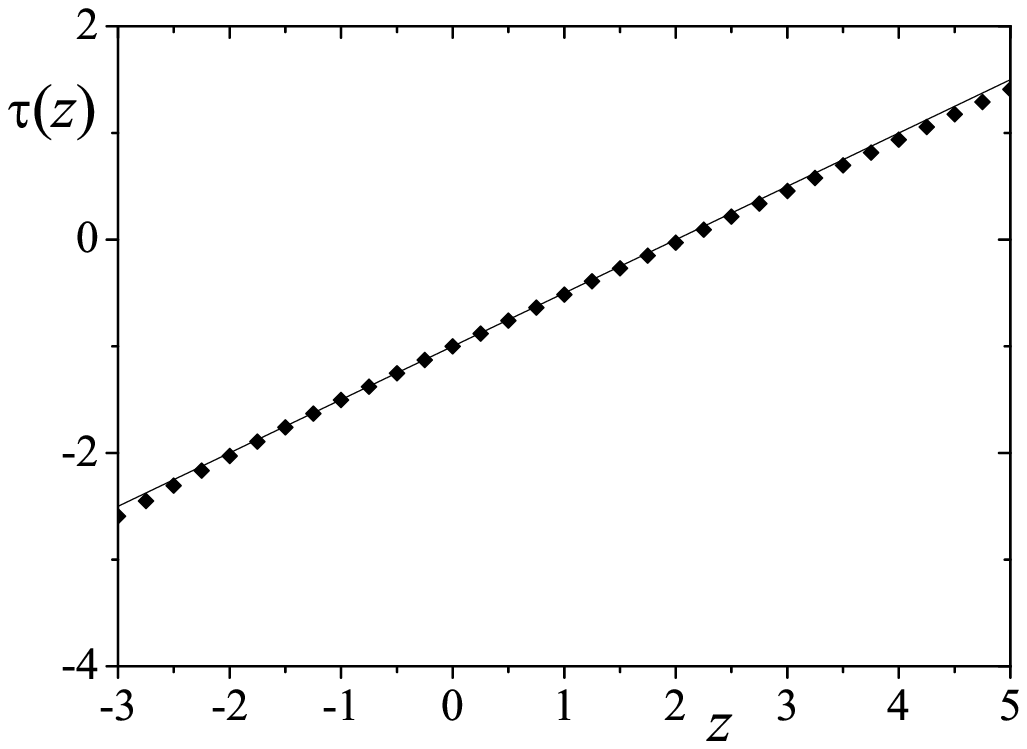} %
\includegraphics[width=0.35\columnwidth,angle=0]{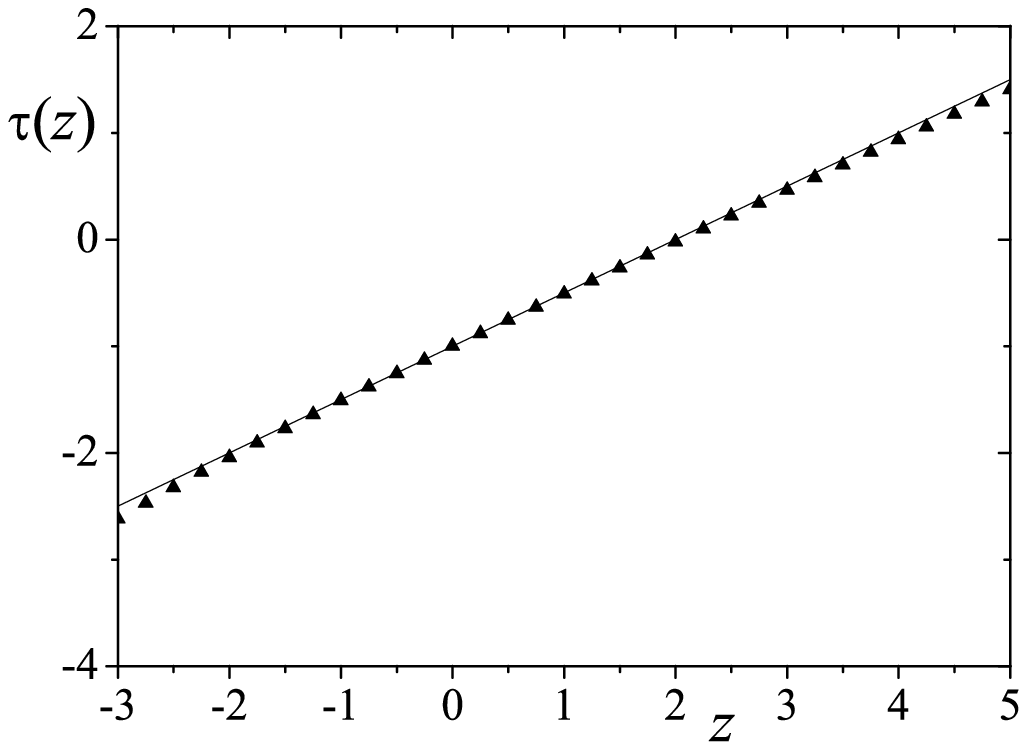}
\caption{ Scaling exponent $\protect\tau $ \textit{vs.} $z$ of the price
fluctuations (upper left), shuffled (upper right), phase randomised (lower
left), and shuffled plus phase randomised (lower right) time series of DJIA
equities. In all panels, the line $\protect\tau =\frac{z}{2}-1$ represents
the theoretical curve for an independent and Gaussian time series. For the
shuffled plus phase randomised time series, the points coincide with the
curve of a Gaussian time series of independent elements. For the other
cases, we verify a departure from the line, proving the multifractal
behaviour of price fluctuations. }
\label{fig-ret-tau}
\end{figure}

In Fig.~\ref{fig-ret-tau} we show over different panels the moment $\tau $
as a function of $z$. We observe that only the shuffled plus phase
randomised signal are in compliance with the theoretical curve, $\tau =z/2-1$%
, of an Gaussian time series of independent elements. 

\subsection{Multifractality for instantaneous volatility time series}

\begin{figure}[th]
\centering 
\includegraphics[width=0.75\columnwidth,angle=0]{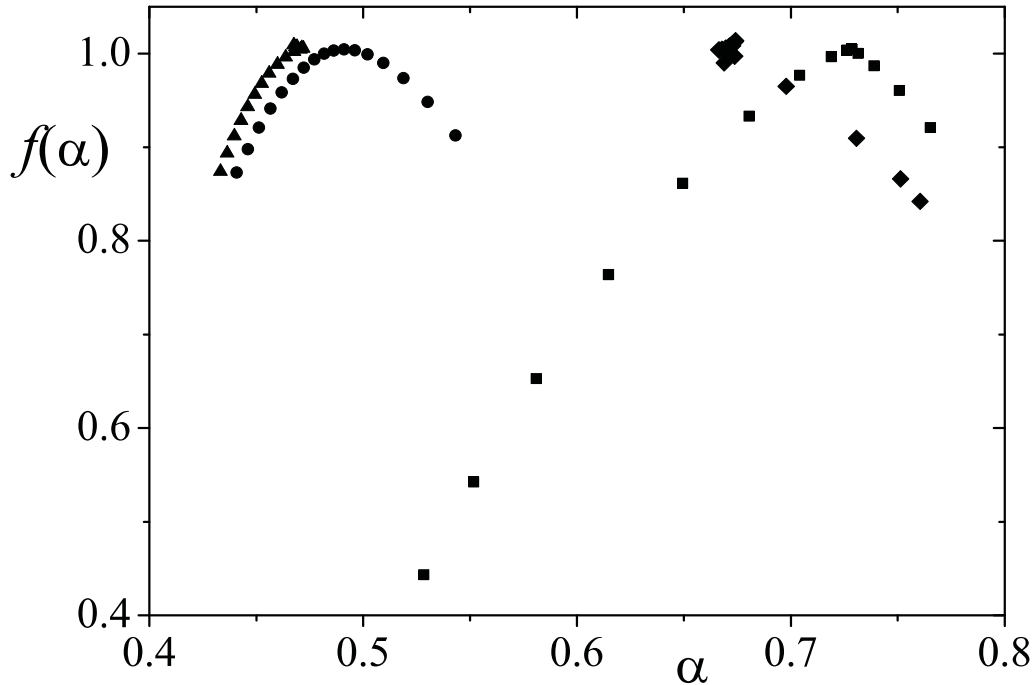}
\caption{ Multifractal spectra $f$ \textit{vs.} $\protect\alpha $ of the
instantaneous volatility ($\blacksquare $), shuffled time series ($\bullet $%
), phase randomised ($\blacklozenge $), and shuffled plus phase randomised ($%
\blacktriangle $) time series of DJIA equities. As it is shown, when
elements that introduce multi-scaling are removed the multi-fractal spectrum
becomes clearly narrower. }
\label{fig-vol-falfa}
\end{figure}

\begin{figure}[th]
\centering
\includegraphics[width=0.35\columnwidth,angle=0]{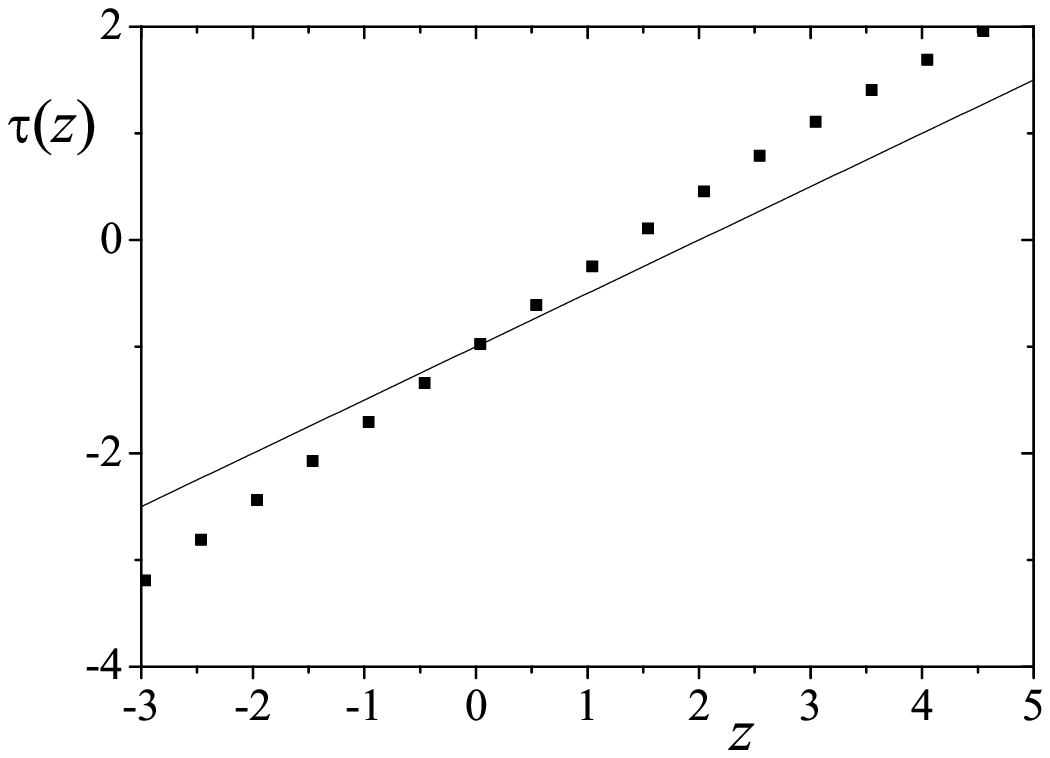} 
\includegraphics[width=0.35\columnwidth,angle=0]{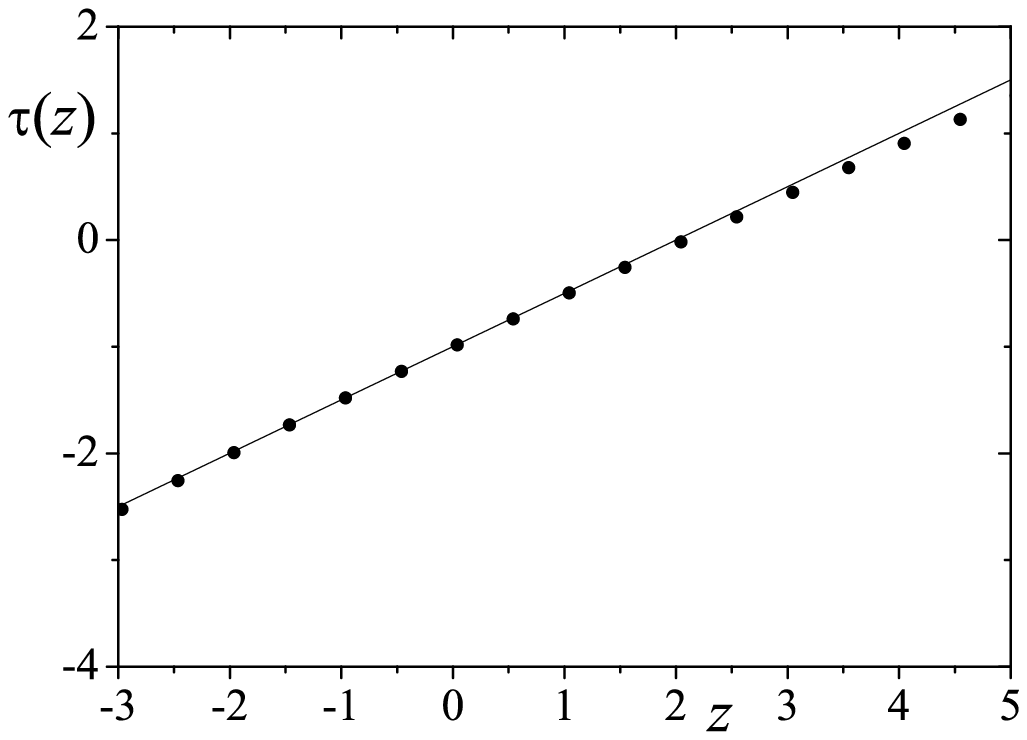} %
\includegraphics[width=0.35\columnwidth,angle=0]{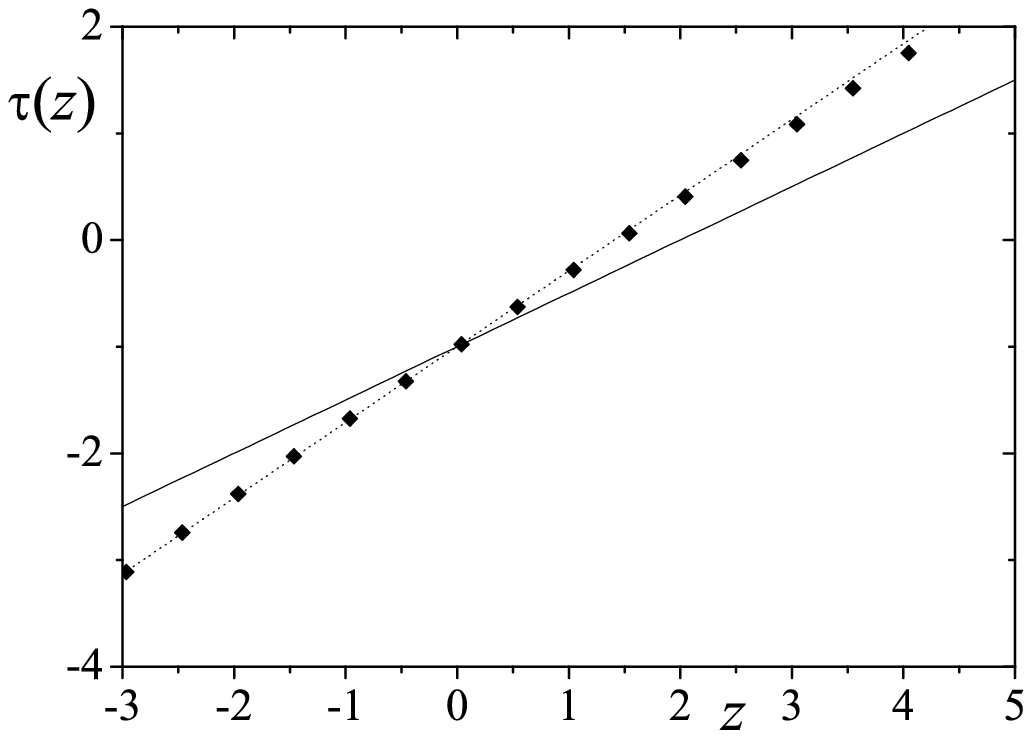} %
\includegraphics[width=0.35\columnwidth,angle=0]{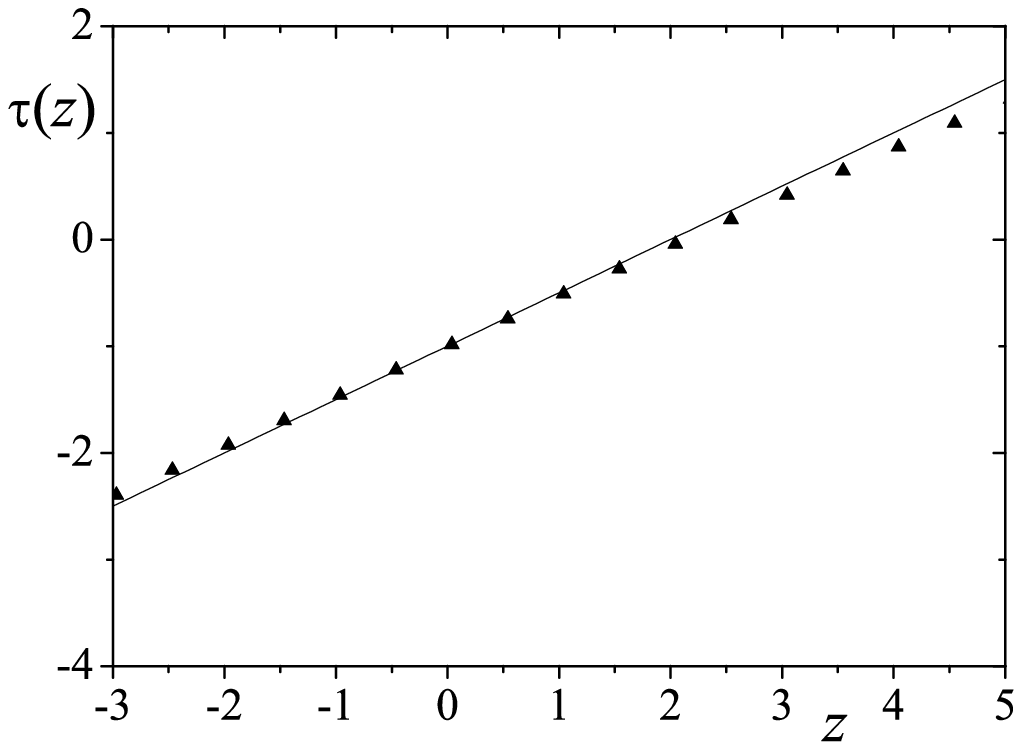} %
\caption{ Scaling exponent $\protect\tau $ \textit{vs.} $z$ of the
instantaneous volatility (upper left), shuffled (upper right), phase
randomised (lower left), and shuffled plus phase randomised (lower right)
time series of DJIA equities. In all panels, the solid line $\protect\tau =\frac{z%
}{2}-1$ represents the theoretical curve of Gaussian time series of
independent elements. Regarding instantaneous volatility results, it is
visible the departure from Gaussian independent behaviour that persists when
we destroy the Gaussianity. In the lower left panel the dotted line represents the 
monofractal curve $\protect\tau = H \, z - 1$ with $H=h(2)=0.71 \pm 0.01$. If we considered the phase randomised time 
series as a pure monofractal set we would have the best fit for $H=0.692 \pm 0.002$, a bit ouside error margin of 
h(2).}
\label{fig-vol-tau}
\end{figure}

Albeit volatility is not directly observable, it plays a central role in
financial modelling \cite{engle-review}, and it is usually related to the
magnitude of price fluctuations. It is on this quantity that long-lasting
covariances associated with asymptotic power-laws are measured. As a matter
of fact, the appropriate mimicry of a long-lasting autocorrelation function
of the volatility associated with a white noise character of the variable
upon study is one of prime challenges in several areas of scientific
research. Aiming to appraise its potential multiscaling nature we have
performed a MF-DFA analysis on instantaneous volatility time series. The
main results are shown in Fig.~\ref{fig-vol-falfa} and Fig.~\ref{fig-vol-tau}%
. From our analysis, we have verified that there are clear differences
between multifractal spectra for price fluctuations and absolute values. 

\begin{figure}[th]
\centering 
\includegraphics[width=0.75\columnwidth,angle=0]{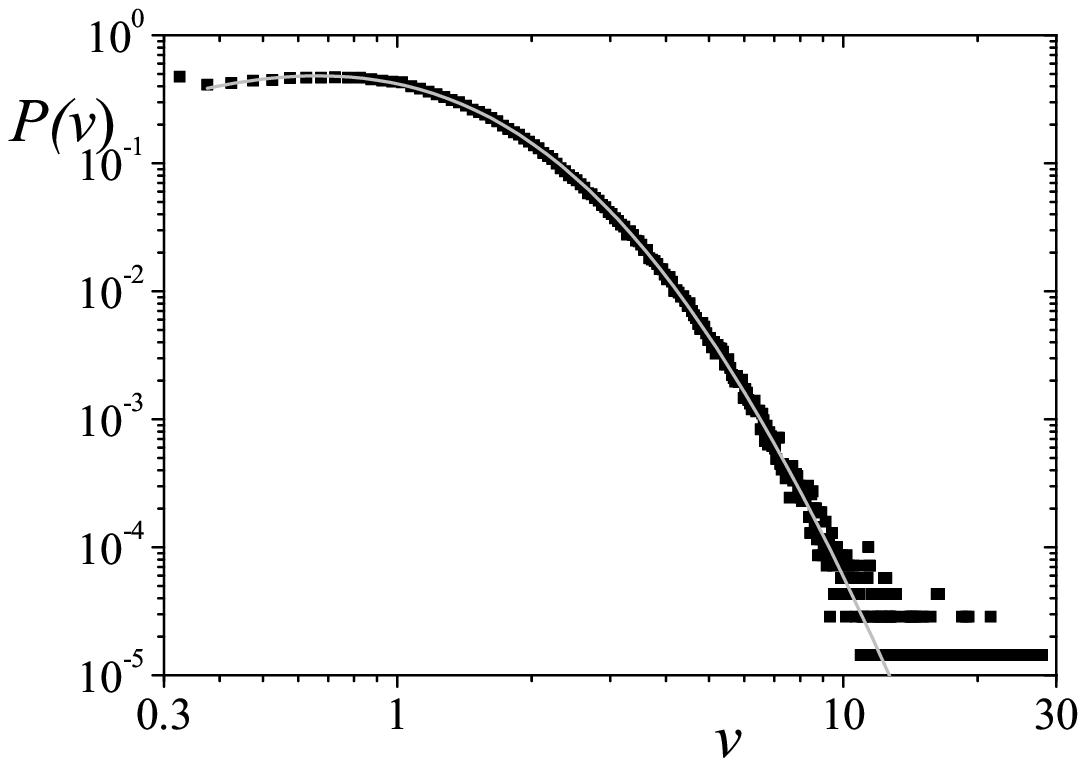}
\caption{ Instantaneous volatility probability density function $P\left(
v\right) $ vs. $v$ averaged over DJIA equities. Symbols are the empirical
PDF and the line the best fit using a $F$-distribution, Eq. (\protect\ref%
{f-dist}) ($\protect\chi ^{2}/n=4.4\times 10^{-6}$ and $R^{2}=0.999$). }
\label{pdf-vol}
\end{figure}

In first place, and against our primary expectations, we have observed that
price fluctuations have a wide multifractal spectrum. Specifically, we have
computed $\Delta h=0.15$ for price fluctuations, and $\Delta h=0.10$ for
volatilities. This corresponds to a ratio of $2$ over $3$. As it happens for
price fluctuations, the multifractal spectrum is asymmetric. We have also
obtained $h\left( 2\right) =0.71$, which indicates a strong persistency on
volatility time series in accordance with previous empirical findings. We
clarify that we expected to obtain a wider spectrum for instantaneous
volatility because of correlations and non-Gaussianity of this quantity. For
shuffled instantaneous volatility time series we observe a shift of $f\left(
\alpha \right) $, and a lessen of curve width. On the other hand, when we
turn instantaneous volatility into a Gaussian variable points multifractal
tends do be clearly diminished, though still present. This is in accordance
to previous verifications about local fluctuations on Hurst exponent for
financial time series \cite{cps-volatility} which introduce multifractality.
Bearing in mind the value $\Delta h=0.10$, the difference between scaling
exponents of the shuffled time series, $\Delta h_{shf}=0.05$, points
non-Gaussianity and dependence as equally responsible for the multiscaling
of instantaneous volatility. From Fig. \ref{fig-vol-tau}, it is visible that 
$\tau _{shf}$ almost coincides with the theoretical curve of an independent
and Gaussian time series. Such a result indicates that the probability
density function presents a nearly exponential decay. We corroborate this
result with Fig.~\ref{pdf-vol} in which we present absolute values
probability density function, $p\left( v\right) $. In line with Fig.~\ref%
{pdf-vol} we verify that $p\left( v\right) $ fits for a $F$-distribution,%
\begin{equation}
F\left( v\right) \varpropto \left( \frac{v}{\theta }\right) ^{\phi }\left[
1-\left( 1-q\right) \frac{v}{\theta }\right] ^{\frac{1}{1-q}},
\label{f-dist}
\end{equation}%
where $\theta =0.32\pm 0.02$, $\phi =1.83\pm 0.01$, and $q=1.08\pm 0.02$.
Taking into account error margins, the small deviation from exponential
decay given by numerical adjustment is in agreement with the slight
deviation of $\tau _{shf}$ from the theoretical curve that we have measured. 

\subsection{Effects of the signal and (instantaneous) volatility
multifractal behaviour on price flctuations multiscaling\label%
{efeitos-volatilidade}}

\begin{figure}[th]
\centering 
\includegraphics[width=0.75\columnwidth,angle=0]{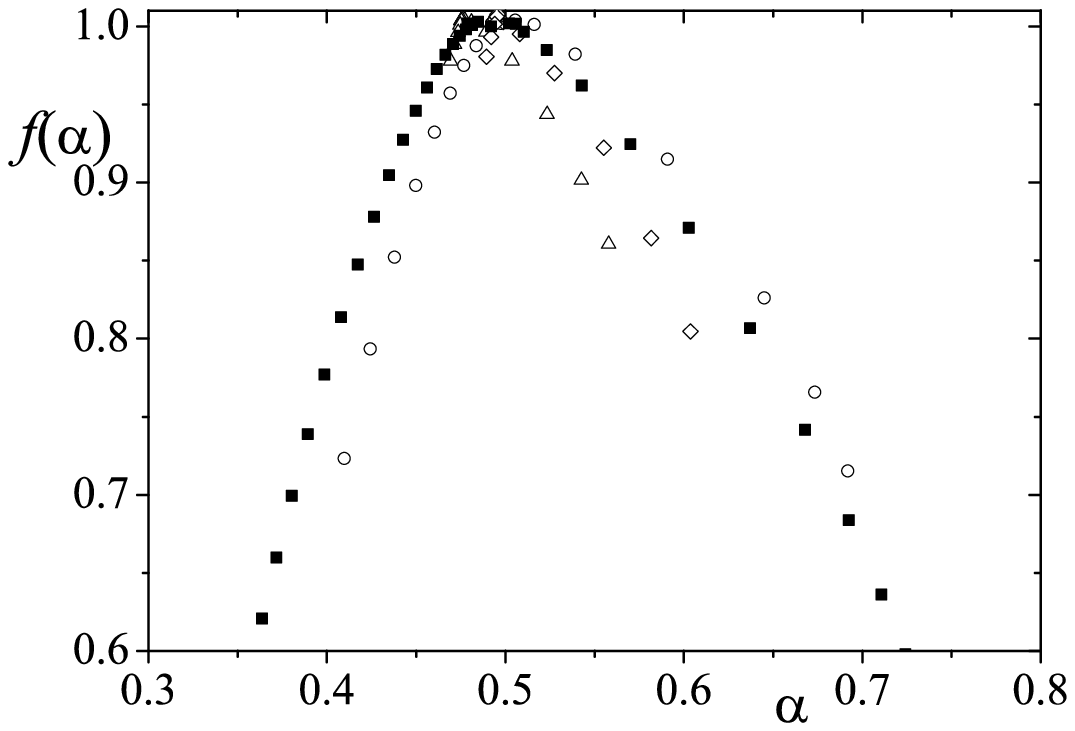}
\caption{Multifractal spectra $f$ \textit{vs.} $\protect\alpha $ of the
price fluctuations ($\blacksquare $), and of time series $\left\{ r\left(
t\right) \right\} $ that use shuffled ($\circ $), phase randomised ($%
\Diamond $), and shuffled plus phase randomised ($\triangle $) volatility
time series of DJIA equities. As it is depicted, the multifractal character
of volatility plays an essential role at the multifractal nature of price
fluctuations. This role is clear for the non-Gaussianity of $v\left(
t\right) $. }
\label{fig-volsig-falfa}
\end{figure}

In this subsection we assess the influence of the multifractal character of
instantaneous volatility on the multifractal nature of price fluctuations.
To that, we have proceeded the following way. We have separated price
fluctuations, $r\left( t\right) $, considering each element as the product
of elements of two other time series, \textit{i.e.}, one that considers the
signal of the price fluctuation, $s\left( t\right) =\pm 1$, and other which
takes into account the magnitude or instantaneous volatility, $v\left(
t\right) =\left\vert r\left( t\right) \right\vert $. Preserving the signal
time series, we have multiplied $\left\{ s\left( t\right) \right\} $ by time
series that were obtained after shuffle, $v_{shf}\left( t\right) $, phase
randomisation, $v_{rnd}\left( t\right) $, and shuffle plus phase
randomisation, $v_{shf-rnd}\left( t\right) $, procedures. The results we
have obtained are depicted in Fig. \ref{fig-volsig-falfa} and Fig. \ref%
{fig-volsig-tau}. From Fig. \ref{fig-volsig-falfa}, we see that the
statistical properties of volatility \textit{do} influence the multifractal
spectrum of price fluctuations. If we only shuffle $\left\{ v\left( t\right)
\right\} $ elements, the $u\left( t\right) $ time series,%
\[
u\left( t\right) \equiv s\left( t\right) \,v\left( t\right) , 
\]%
just has a paltry narrower $f\left( \alpha \right) $ curve than $\left\{
r\left( t\right) \right\} $. It has $\Delta h=0.13$ in opposition to $\Delta
h=0.15$ of $\left\{ r\left( t\right) \right\} $. This is an unexpected
result regarding the influence of $\left\{ v\left( t\right) \right\} $
ordering on its multifractal spectrum. However, when we destroy the
non-Gaussianity of instantaneous volatility probability density function, we
basically destroy the multifractal spectrum of price fluctuations, since $%
\Delta h=0.04$, or $\Delta h=0.03$ when we combine shuffling with phase
randomisation procedures on $\left\{ v\left( t\right) \right\} $. The latter
result also sets the influence of the signal ordering on the price
fluctuations multifractal character at the order of error in absolute
accordance with previous analysis for other characteristics, namely the
approach to the Gaussian when of cummulative price fluctuations probability
density functions \cite{lyra}. 
\begin{figure}[th]
\centering 
\includegraphics[width=0.31\columnwidth,angle=0]{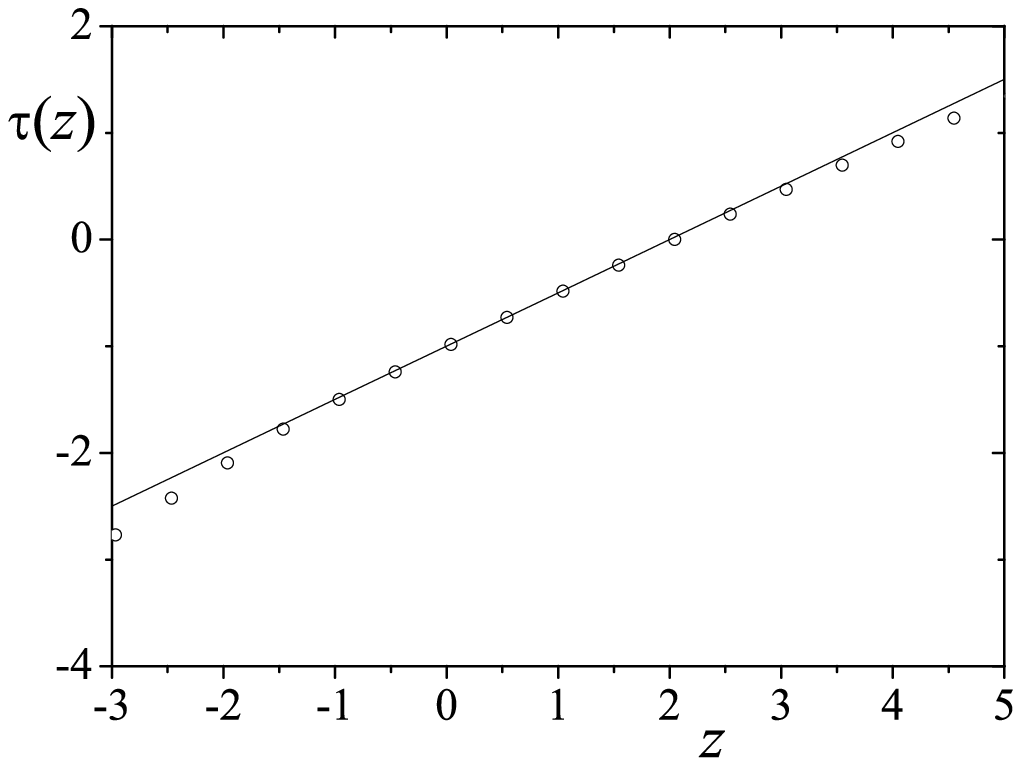} %
\includegraphics[width=0.31\columnwidth,angle=0]{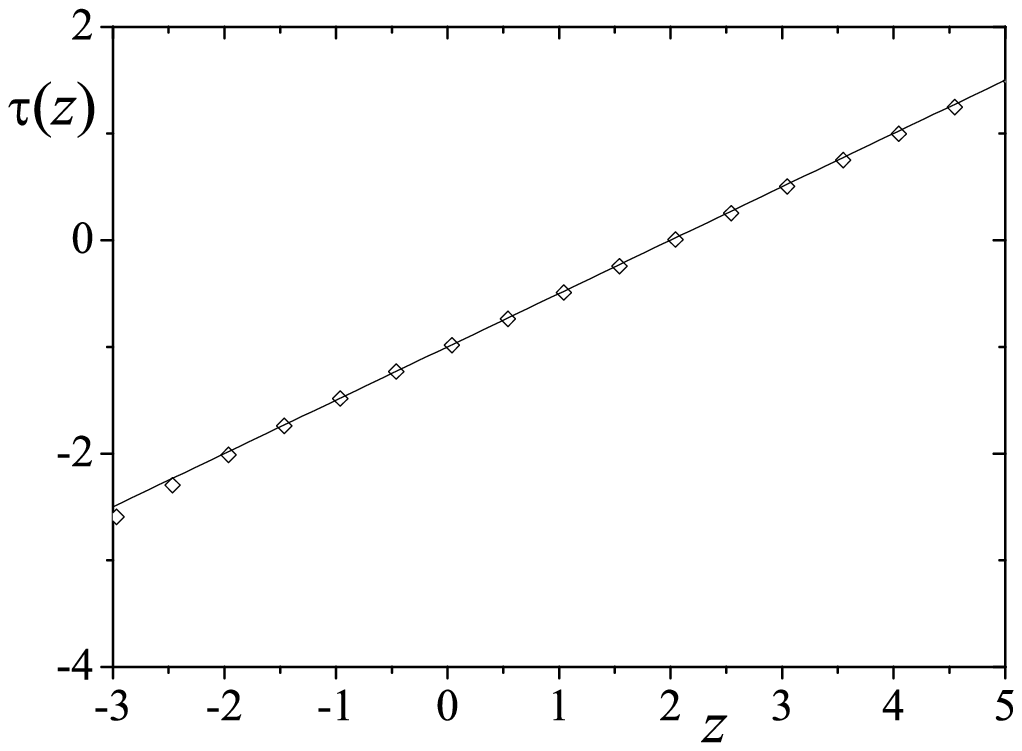} %
\includegraphics[width=0.31\columnwidth,angle=0]{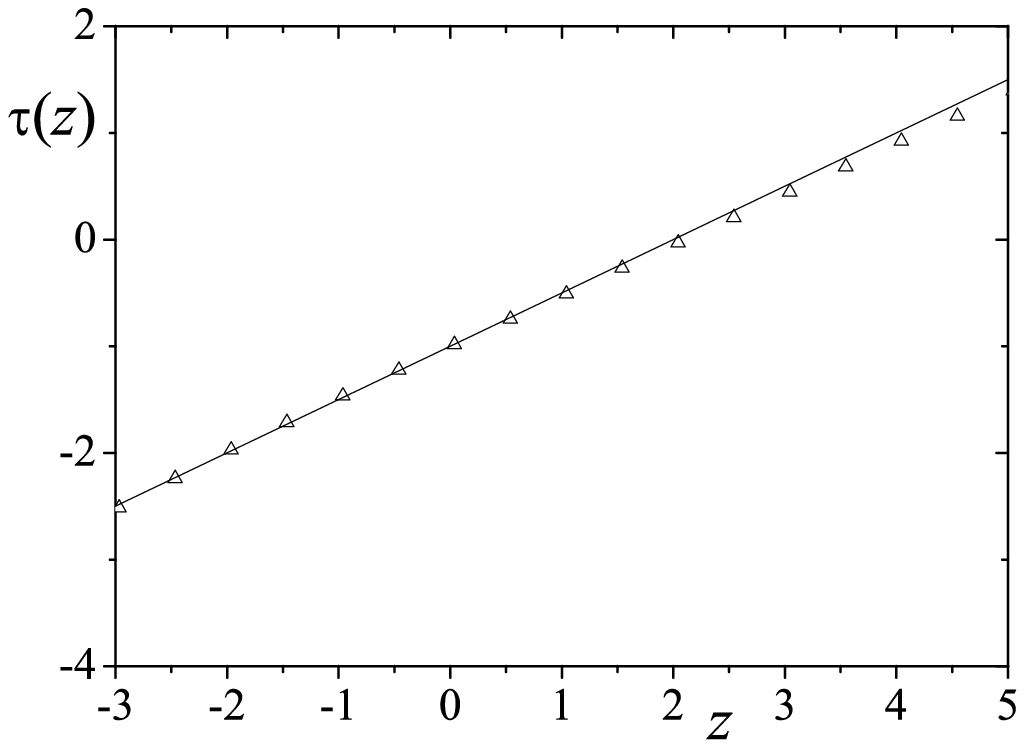}
\caption{Scaling exponent $\protect\tau $ \textit{vs.} $z$ of time series $%
\left\{ r\left( t\right) \right\} $ that use shuffled (right), phase
randomised (centre), and shuffled plus phase randomised (left) volatility
time series of DJIA equities. In all panels, the line $\protect\tau =\frac{z%
}{2}-1$ represents the theoretical curve for an independent and Gaussian
time series. The importance of the multifractal characteristics of
volatility are demonstrated by the clear approach of these three results
towards the theoretical curve of an independent Gaussian process. }
\label{fig-volsig-tau}
\end{figure}

As it has been observed \cite{bouchaud}\cite{smdq-qf}, many of the dynamical
and statistical properties of price fluctuations depend on the volatility.
Although it is a pivotal variable in finance the truth is that volatility
definition is still ambiguous \cite{engle-gallo}. If in many situations it
is presented has we have been doing, volatility is oftenly determined as the
standard deviation of price fluctuations over window of length $l$ \footnote{%
When $l=1$ we obtain the instantaneous volatility definition.}. The latter
definition is widely applied on stochastic volatility models. In that
particular case, superstatistical models have been applied in problems of
financial origin \cite{bouchaud}\cite{sato} to define such models.
Concisely, \emph{superstatistics} or \emph{\textquotedblleft statistics of
statistics\textquotedblright } \cite{beck-cohen} is a compound method which
has emerged within statistical mechanics. It is based on the assumption of a
local statistics dependent on a parameter that fluctuates (smoothly) on a
time scale that is very large when compared with the time needed for a
system to reach a local equilibrium or stationarity. In a superstatistical
context it has been proved that, if we have a set of local Gaussian random
variables,%
\begin{equation}
\mathcal{G}_{\sigma }\left( x\right) =\frac{1}{\sqrt{2\pi }\sigma }\exp %
\left[ -\frac{x^{2}}{2\,\sigma ^{2}}\right] ,  \label{gauss}
\end{equation}%
and the inverse variance, $\sigma ^{-2}$, is associated with a $\Gamma $%
-distribution,%
\begin{equation}
\Gamma \left( x\right) \varpropto \left( \frac{x}{\delta }\right) ^{\gamma
}\exp \left[ -\frac{x}{\delta }\right] ,  \label{g-dist}
\end{equation}%
then, the stationary distribution given by%
\[
p\left( x\right) =\int \mathcal{G}_{\sigma }\left( x\right) \,\Gamma \left(
\sigma ^{-2}\right) \,d\left( \sigma ^{-2}\right) , 
\]%
is equal to a Tsallis (or Student $t$-) distribution \cite{tsallis-milan},%
\begin{equation}
p\left( x\right) =\frac{1}{Z}\left[ 1-\left( 1-q\right) \,\frac{x^{2}}{%
\lambda }\right] ^{\frac{1}{1-q}},  \label{q-gauss}
\end{equation}%
where 
\begin{equation}
q=1+\frac{2}{3+2\,\gamma }.  \label{q-superstat}
\end{equation}%
In this way, superstatistics has been considered has the first dynamical
foundation for non-extensive framework \cite{beck-prl} that has non-additive
entropy, $S_{q}$ \cite{ct}, as its cornerstone. Distribution (\ref{q-gauss})
has regularly been used to fit for price fluctuations of several financial
markets, and also for the data set we have been analysing for which it has
been found a value of $q=1.31\pm 0.02$ \cite{canberra}. If we assume a
superstatistical approach for the data set upon analysis from Eq. (\ref%
{q-superstat}) we obtain $\gamma =1.82$.

\begin{figure}[th]
\centering 
\includegraphics[width=0.45\columnwidth,angle=0]{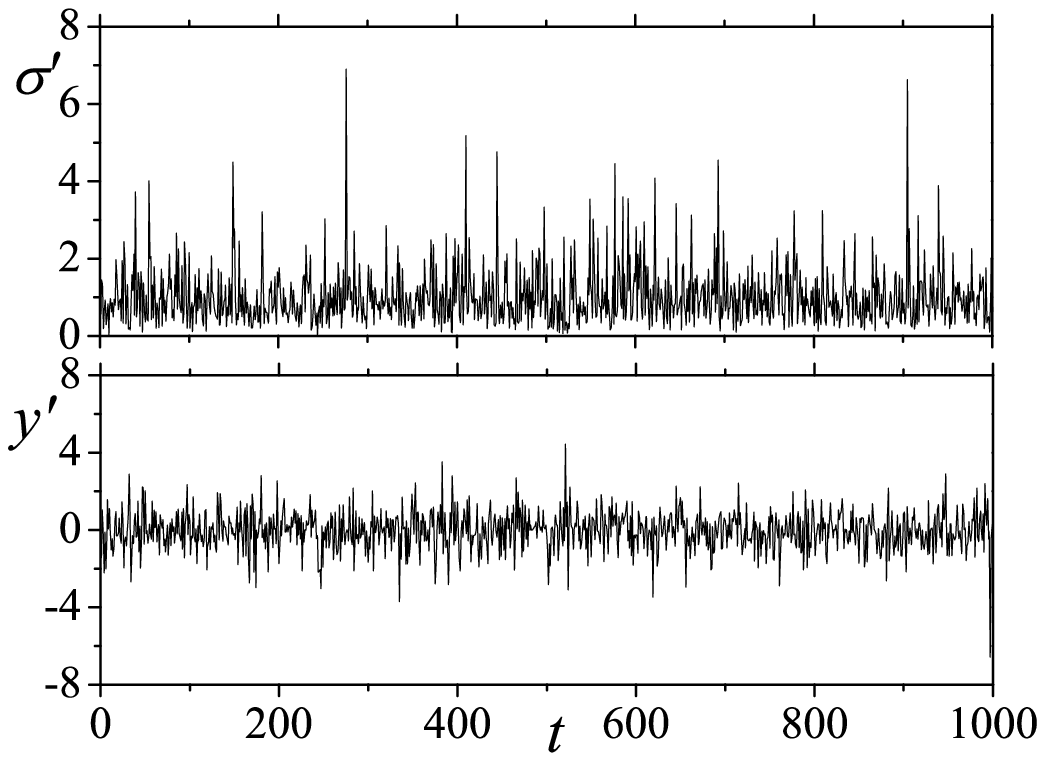} %
\includegraphics[width=0.45\columnwidth,angle=0]{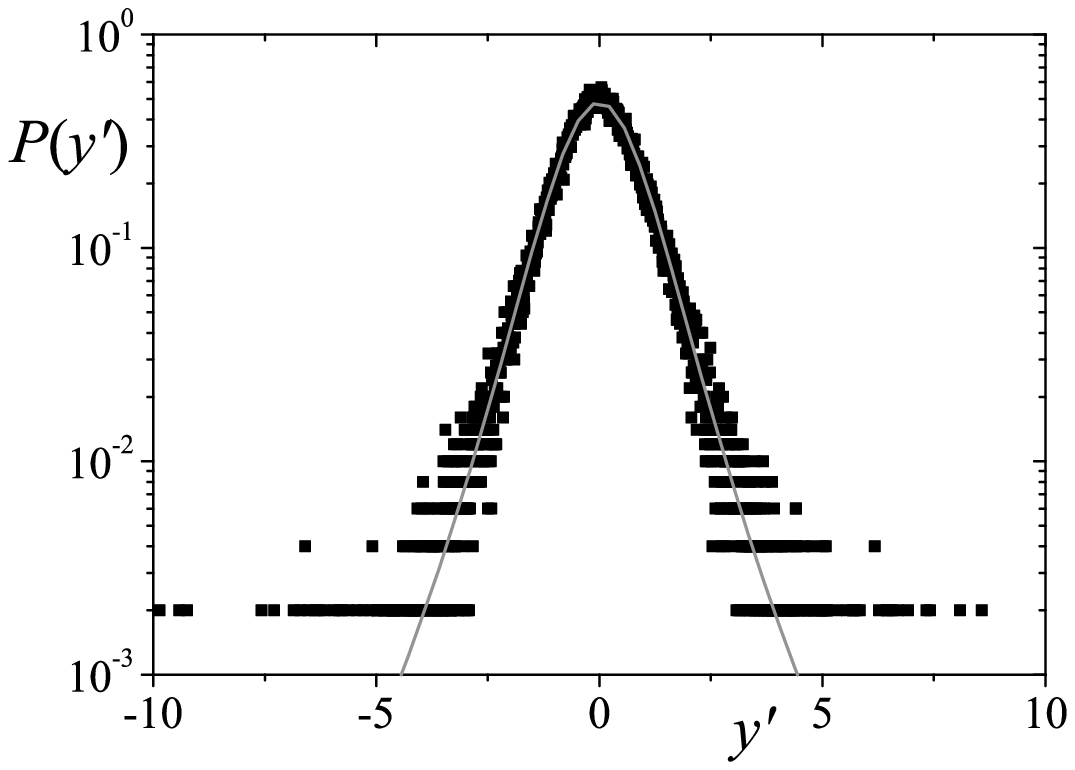}
\caption{ Left: Excerpt of the $ARCH$-like signal (lower panel) where $%
y^{\prime }$ represents $y$ divided by the average value of $\protect\sigma $%
, $\protect\sigma _{m}$. The elements of $\protect\sigma $ signal (upper
panel) follow a PDF such that $P\left( \protect\sigma ^{-2}\right) $ is a $%
\Gamma $-distribution with $\protect\gamma =1.82$, and $\protect\delta =2$.
Right: Stationary PDF of $y^{\prime }$, $P\left( y^{\prime }\right) $ vs. $%
y^{\prime }$. Symbols have been obtained from the time series and the line
is the numerical adjustment for a $q$-Gaussian distribution with $q=1.3$.
Although this is not an exact approach, the adjustment is rather nice ($%
\protect\chi ^{2}/n=1.2\times 10^{-5}$ and $R^{2}=0.99$). }
\label{return}
\end{figure}

In what follows we analyse a discrete $ARCH$-like process~\cite{engle-arch}
that can be catalogued as superstatistical. Explicitly, we have generated
time series, $\left\{ y\left( t\right) \right\} $, from the product of an
uncorrelated Gaussian signal, $\left\{ \omega \left( t\right) \right\} $,
with $\left\langle \omega \left( t\right) \right\rangle =0$, and $%
\left\langle \omega \left( t\right) ^{2}\right\rangle =1$ by an uncorrelated
volatility signal, $\left\{ \sigma \left( t\right) \right\} $,%
\[
y\left( t\right) \equiv \sigma \left( t\right) \,\omega \left( t\right) , 
\]%
such that $\sigma ^{-2}$ follows a $\Gamma $-distribution with $\gamma =1.82$
as we have obtained. In this case, because we neglect memory on volatility,
we can compare the multifractal study of this time series with the results
that we have presented at the beginning of this subsection \ref%
{efeitos-volatilidade} for $\left\{ u\left( t\right) \right\} $ with a
shuffled instantaneous volatility. We have opted for this comparison
because, just as $s\left( t\right) $, $\omega \left( t\right) $ does not
contribute to the multifractal spectrum. The excerpt of the time series we
have generated is presented in Fig.~\ref{return}. In the same figure we
comprove that $\left\{ y\left( t\right) \right\} $ follows PDF (\ref{q-gauss}%
) with $q=1.3$.

\begin{figure}[th]
\centering\includegraphics[width=0.45\columnwidth,angle=0]{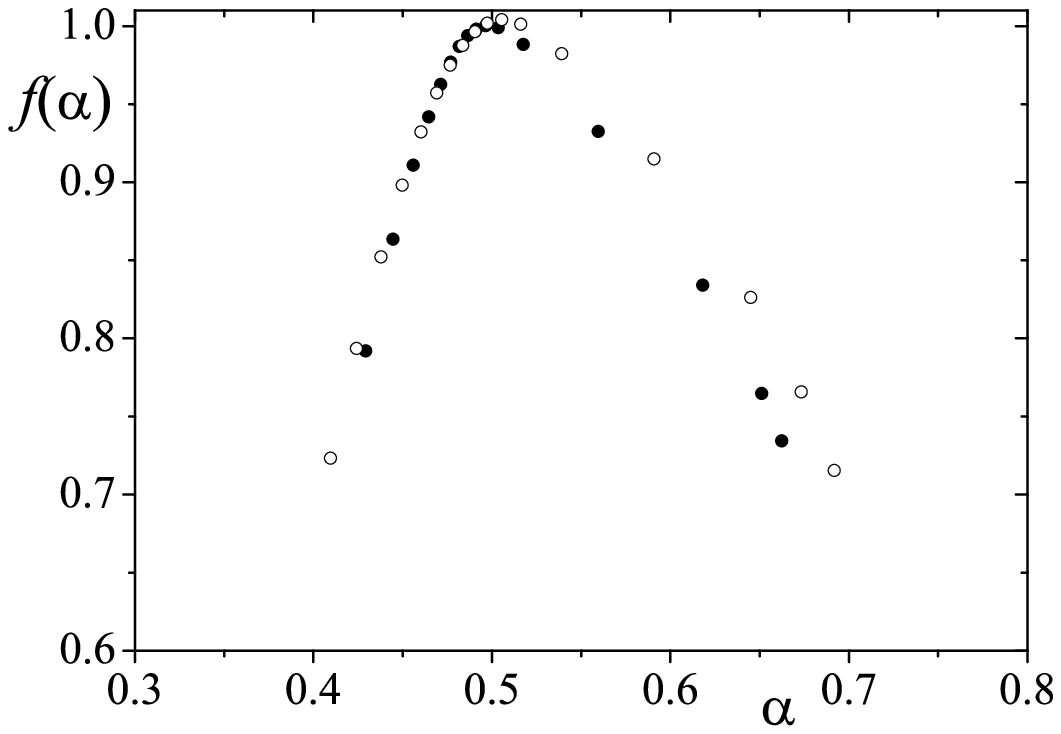}%
\includegraphics[width=0.45\columnwidth,angle=0]{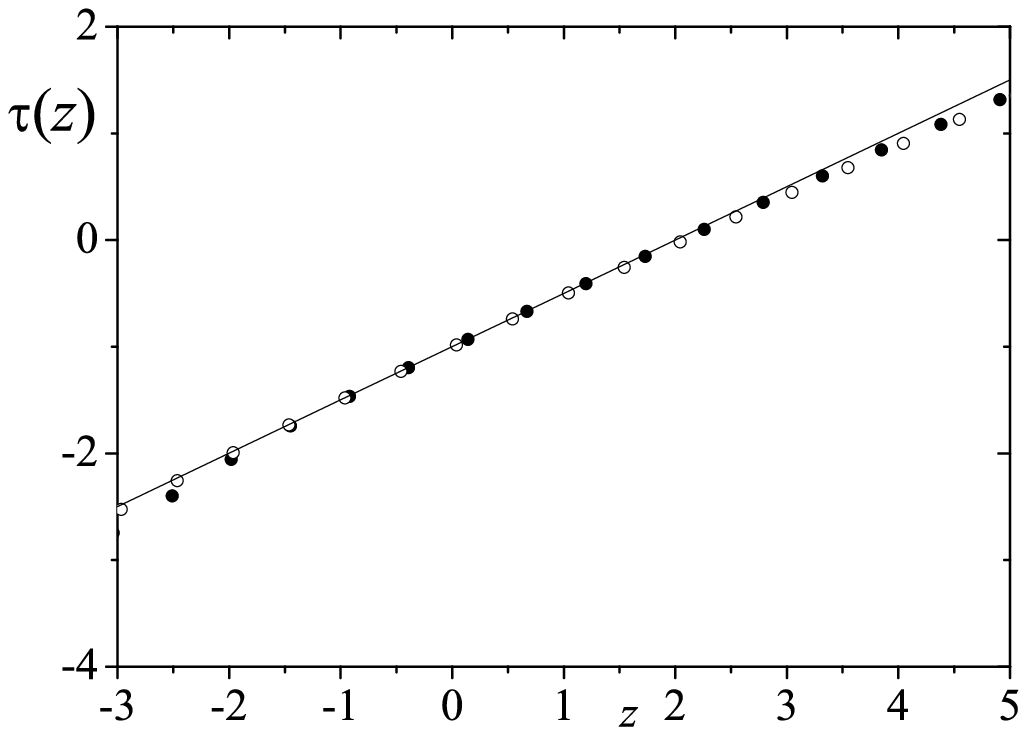}
\caption{Left panel: Multifractal spectra $f$ \textit{vs.} $\protect\alpha $
of the $\left\{ y\left( t\right) \right\} $ ($\bullet $), and of time series 
$\left\{ r\left( t\right) \right\} $ that use shuffled ($\circ $) volatility
time series of DJIA equities. Right panel: Scaling exponent $\protect\tau $ 
\textit{vs.} $z$ of $\left\{ y\left( t\right) \right\} $ ($\bullet $) and of
time series $\left\{ r\left( t\right) \right\} $ that use shuffled ($\circ $%
) volatility time series of DJIA equities. The line $\protect\tau =\frac{z}{2%
}-1$ represents the theoretical curve for an independent and Gaussian time
series.}
\label{fig-superstat}
\end{figure}

Afterwards, we have performed a multifractal analysis along the same lines
we have made for price fluctuations. Even though both multifractal spectra
are very similar we can verify a noticeable difference. As a matter of fact
we have obtained $\Delta \alpha =0.28$ for $\left\{ y\left( t\right)
\right\} $ with shuffled volatility time series, and $\Delta \alpha =0.24$
for the generated series that we have priory analysed, \textit{i.e.}, an
error of $17\%$, see in Fig. \ref{fig-superstat}. This means that
superstatistics can be considered as an acceptable first approach, although
models that consider long term memory in variance \cite{qarch} are certainly
more appropriate. Since the only source of multifractality in this case is
the asymptotic power-law behaviour of the stationary PDF $P\left( y\right) $%
, $\left\{ y\left( t\right) \right\} $ should be in fact called a \textit{%
bifractal} with $\tau \left( z\right) =0$ for $z>5.45$.

\section{Results for $\ell $-diagrams~\label{l-diagrams}}

\begin{figure}[tbp]
\includegraphics[width=0.75\columnwidth,angle=0]{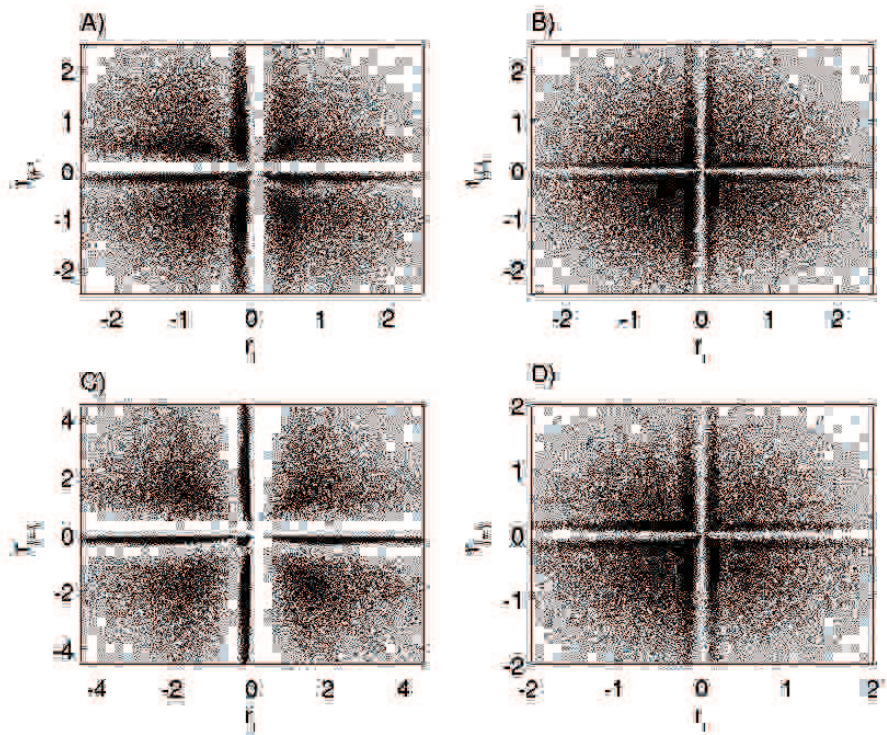}
\caption{Recurrence maps (step $s=1$) for the companies Caterpilar (A),
Citibank (B), Intel (C) and 3M (D) - detrended data.($i \equiv t $)}
\label{fig-ret-map}
\end{figure}

A rich and interesting way of representing time series is to consider a
mapping of the time series onto a plane where each point signalled is
obtained by pairing elements $x_{t}$ and $x_{t+\ell }$ of the time series as
ordinate and abcissa, respectively. This $\ell $-diagrams and related
methods~\cite{rp} are frequently used on studies about biological \cite{ldv}%
, and dynamical systems \cite{rp}. Moreover, \ they have also been
introduced to study daily fluctuations of some securities \cite%
{ausloos-return}. This type of representation, full called as $\ell $\textit{%
-diagram variability method} \cite{ldv}, is in fact quite illustrative since
it is a simple way of capturing regular aspects of systems which are
apparently irregular. Such regularities can be characterised by regions
which are more visited in space $x_{t}\times x_{t+\ell }$. Specifically,
taking into account price fluctuations time series and $\ell =1$ as an
example, it allows one to verify how prices evolve in segments of two time
intervals. Nextly, we analyse the first return map of the price
fluctuations. In Fig. \ref{fig-ret-map} we show the plot of $r_{t+1}$ 
\textit{versus} $r_{t}$ for some of the companies of our set \footnote{%
The plotted companies have been chosen in order to represent different
sectors of activity and ways of trading (NYSE and NASDAQ).}. The plots
present a very interesting structure. Over the four quadrants
(anticlockwise) we have got stripes with high density of points and
\textquotedblleft forbidden\textquotedblright\ regions close to the axes. We
assign to transaction costs the emergence of this banned regions. In the $%
3^{rd}$ quadrant we can see a highly visited region close to the origin,
point that small decreases induce small decreases. We have investigated the
probabilities for each quadrant and we have found a very peculiar behaviour
for DJ30 \footnote{%
We have also calculated these probabilities for original data - intra-day
trend mask the effects observed in first return maps - and, once again, we
have found the same behaviour, but with different probabilities.}. The
probabilities for each quadrant (anticlockwise) can be interpreted as
follows: $1^{st}$ quadrant - probability of two consecutive profits, $P_{1}$%
; $2^{nd}$ quadrant - probability of a profit after a loss, $P_{2}$; $3^{rd}$
quadrant - probability of two consecutive loss, $P_{3}$; $4^{th}$ quadrant -
probability of a loss after profit, $P_{4}$. These results are shown in
Table \ref{tabela}. As it is easily observable the dynamics of the system is
basically up-and-down-and-up since the fourth and second quadrants together
represent $2/3$ of the points plotted on 1-diagrams, with the probabilities
of having either two consecutive profits or two consecutive losses equal to $%
17\%$, in average. Interestingly, we have verified that although the number
of negative price fluctuations surpasses the number of positive price
fluctuations, $N\left( r_{i}>0\right) -N\left( r_{i}<0\right) =-199$
(related to the skewness of the distributions), the cumulative sum of price
fluctuations yields a positive value for all equities, $\sum_{t}r_{i}\left(
t\right) =3.82$ (in average). In other words, although during the period
upon analysis there was a larger number of negative price fluctuations than
positive price fluctuations, the magnitude of the latter were greater so
that a positive evolution arose. As a matter of fact, during this period the
DJIA index increased its magnitude from $10334.16$ to $10783.01$, or a
heighten of $4.8\%$.

\begin{figure}[tbp]
\includegraphics[width=0.75\columnwidth,angle=0]{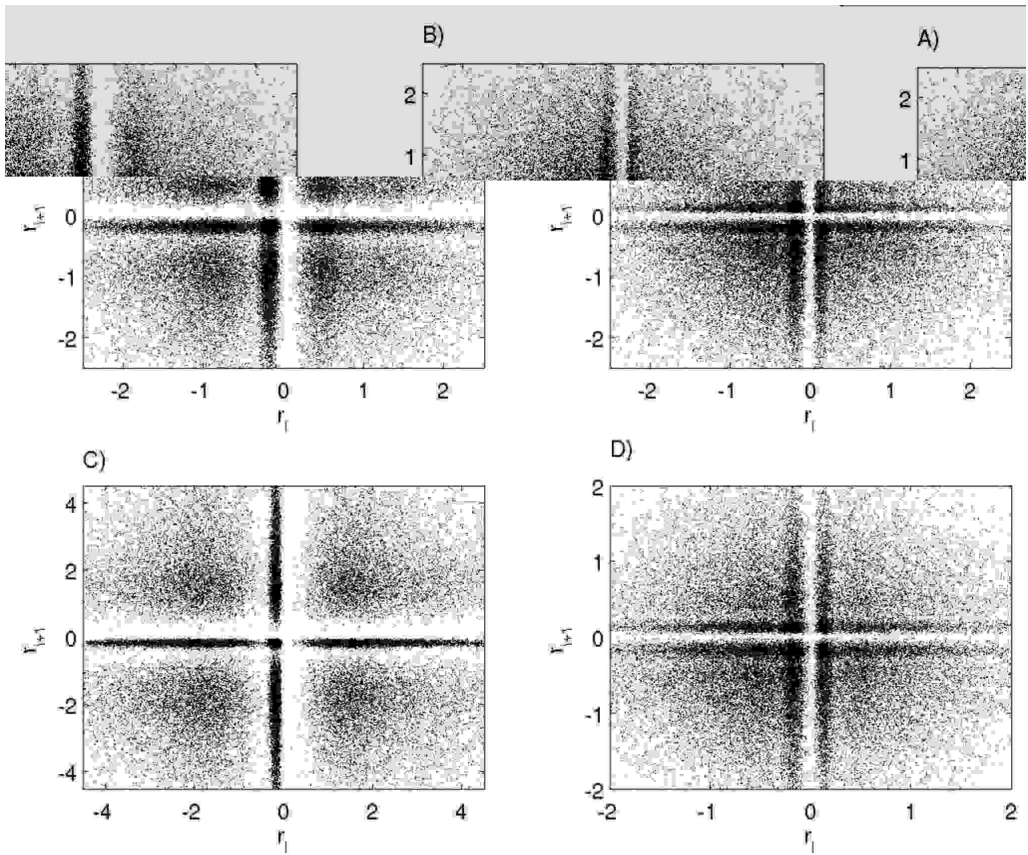}
\caption{Recurrence maps (step $s=1$) for the companies Caterpilar (A),
Citibank (B), Intel (C) and 3M (D) - detrended and shuffled data.($i \equiv
t $)}
\label{fig-ret-map-shf}
\end{figure}

\begin{figure}[tbp]
\includegraphics[width=0.75\columnwidth,angle=0]{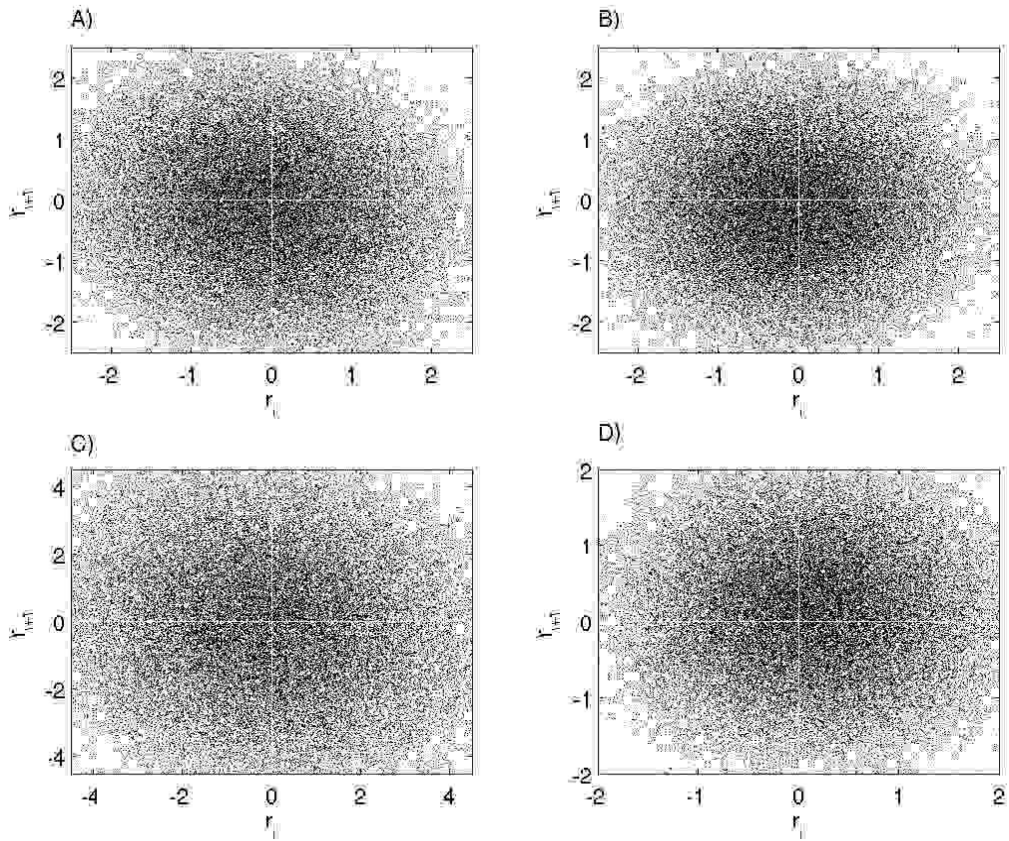}
\caption{Recurrence maps (step $s=1$) for the companies Caterpilar (A),
Citibank (B), Intel (C) and 3M (D) - detrended, shuffled \textit{plus} phase
randomised data.($i \equiv t $)}
\label{fig-ret-map-rnd}
\end{figure}

Be aware that, looking at Fig. \ref{fig-ret-map}, there exists a clear
pattern for these probabilities. In order to further show that these
characteristic patterns go beyond the uncorrelated essence of price
fluctuations time series, we have performed immediate $1$-diagrams for the
shuffled signals. The results are presented in Fig. \ref{fig-ret-map-shf}
where it is visible that these diagrams are different from the diagrams that
we have shown in Fig. \ref{fig-ret-map}, namely the accumulation around
lines $r_{t+1}=\pm r_{t}$ becomes less clear. Furthermore, analysing
shuffled plus randomised times series, Fig. \ref{fig-ret-map-rnd}, we have
observed the lost of any pattern, forbidden stripes inclusive. Actually,
both of the two latter representations are more homogeneous. In our opinion
this is a clear evidence about the importance of dependencies and
non-Gaussianity on price fluctuations dynamics. At this point it is
absolutely necessary to stress that this profile for $1$-diagrams does not
contradics the EMH, if one tried to make use of this property for immediate
trading, transaction costs would surpass any possible (read likely) income.

\begin{table}[tbp]
\caption{Probabilities for each quadrant (columns 2-4) of 30 companies of
the DJ30. In column 6 is shown the difference between
positives price fluctuations and negative price fluctuations, and in column
7 is shown the sum of all returns. The last two columns have been obtained
from trended data. Even though most price fluctuations are negative (for
most companies) though the sum is positive. Note also that there is a clear
pattern for the quadrants for the 30 companies.}\centering
\begin{tabular}{lcccccc}
\hline
\ \ \ \ \ \ \ \ \ \ \ \  & $P_{1} \quad $ & $P_{2} \quad $ & $P_{3} \quad $
& $P_{4} \quad $ & $N\left( r_{i}>0\right) -N\left( r_{i}<0\right) \quad $ & 
$\sum r_{i}$ \\ \hline
aa & 0.17 & 0.33 & 0.17 & 0.33 & -269 & 4.99 \\ 
aig & 0.18 & 0.33 & 0.17 & 0.33 & -55 & 4.16 \\ 
axp & 0.18 & 0.33 & 0.17 & 0.33 & -157 & 1.38 \\ 
ba & 0.18 & 0.33 & 0.17 & 0.33 & -262 & 4.01 \\ 
c & 0.18 & 0.33 & 0.16 & 0.33 & 132 & 4.80 \\ 
cat & 0.17 & 0.33 & 0.18 & 0.33 & -477 & 6.55 \\ 
dd & 0.18 & 0.33 & 0.17 & 0.33 & -130 & 5.09 \\ 
dis & 0.17 & 0.33 & 0.17 & 0.33 & -237 & 5.02 \\ 
ge & 0.17 & 0.33 & 0.17 & 0.33 & -280 & 4.71 \\ 
gm & 0.18 & 0.33 & 0.17 & 0.33 & -366 & 2.92 \\ 
hd & 0.18 & 0.33 & 0.17 & 0.33 & -113 & 0.76 \\ 
hon & 0.18 & 0.33 & 0.17 & 0.33 & -269 & 1.11 \\ 
hpq & 0.17 & 0.33 & 0.17 & 0.33 & 10 & 5.34 \\ 
ibm & 0.18 & 0.32 & 0.17 & 0.32 & -45 & 0.90 \\ 
intc & 0.19 & 0.32 & 0.18 & 0.32 & 172 & 2.23 \\ 
jnj & 0.17 & 0.33 & 0.17 & 0.33 & -96 & 0.57 \\ 
jpm & 0.17 & 0.33 & 0.17 & 0.33 & -276 & 0.76 \\ 
ko & 0.18 & 0.33 & 0.17 & 0.33 & -289 & 6.22 \\ 
mcd & 0.18 & 0.33 & 0.16 & 0.33 & 33 & 1.15 \\ 
mmm & 0.18 & 0.33 & 0.16 & 0.33 & -348 & 7.23 \\ 
mo & 0.18 & 0.33 & 0.16 & 0.33 & -82 & 6.17 \\ 
mrk & 0.17 & 0.33 & 0.17 & 0.33 & -298 & 1.16 \\ 
msft & 0.19 & 0.31 & 0.18 & 0.32 & -16 & 1.58 \\ 
pfe & 0.17 & 0.33 & 0.17 & 0.32 & -162 & 3.63 \\ 
pgn & 0.18 & 0.33 & 0.17 & 0.33 & -62 & 5.16 \\ 
sbc & 0.18 & 0.32 & 0.17 & 0.33 & -336 & 6.54 \\ 
utx & 0.18 & 0.32 & 0.17 & 0.33 & -478 & 5.85 \\ 
vz & 0.17 & 0.33 & 0.17 & 0.33 & -191 & 4.65 \\ 
wmt & 0.17 & 0.33 & 0.18 & 0.32 & -583 & 4.03 \\ 
xom & 0.17 & 0.33 & 0.17 & 0.33 & -452 & 6.02 \\ 
\textbf{average} & \textbf{0.17} & \textbf{0.33} & \textbf{0.17} & \textbf{%
0.33} & \textbf{-199} & \textbf{3.82} \\ \hline
\end{tabular}%
\label{tabela}
\end{table}

Analysing $\ell $-diagrams for $\ell =2,4,10$ we have verified an equal
occupancy of all quadrants,$P_{1}=P_{2}=P_{3}=P_{4}=25\% ,$ which indicates
the loss of any predictability on the time series.

\begin{figure}[tbp]
\includegraphics[width=0.75\columnwidth,angle=0]{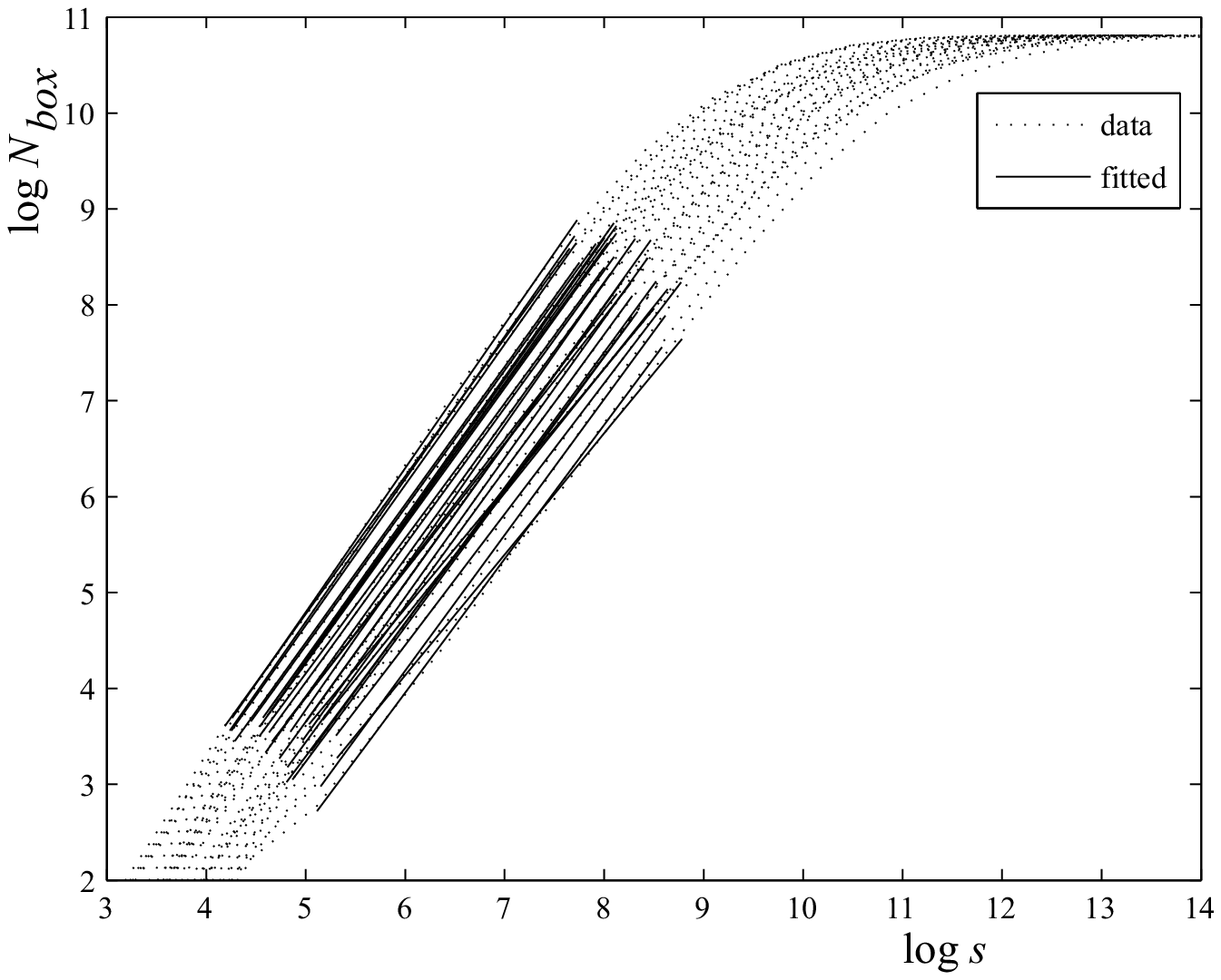}
\caption{Estimation of the fractal dimension of the $r_{t}\times r_{t+1}$
space using Eq. (\protect\ref{boxcount}).}
\label{df_map_ret}
\end{figure}

To quantify the properties of $\ell $-diagrams we have also applied the
algorithm described in Sec. \ref{hou-method}. Namely, we have mapped the
space $r_{t}\times r_{t+1}$ onto interval $[2^{0},2^{16}-1]$, and we have
estimated the fractal dimension of this space structure for the 30
companies. In Fig. \ref{df_map_ret} it is noticeable that for the majority
of companies the scale regime holds over a large range of scales. We have
then used the interval $2^{2}$ to $2^{8}$ to numerically obtain the fractal
dimensions that are shown in Table \ref{tab_df}, which correspond to the
slopes of the fitting straight lines in Fig. \ref{df_map_ret}. Therein, it
is verifiable that the fractal dimension vary slight as the step ($s$) is
changed, as well as after a shuffling procedure (keeping the order of the
diagram). However, it is strongly affected by phase randomization, and it
presents for this case values that are compatible with a 2-dimensional
Gaussian distribution.

\begin{table}[tbp]
\caption{Fractal dimension of the $r_t\times r_{t+ \ell}$ space ($%
\ell=1,2,10,50$) for the companies of the DJ30 estimated with Hou algorithm.}
\label{tab_df}\centering
\begin{tabular}{ccccccc}
\hline
\ \ \ \ \ \ \ \ \ \ \ \  & $d_{f}\left( \ell =1\right) \quad $ & $%
d_{f}\left( \ell =2\right) \quad $ & $d_{f}\left( \ell =10\right) \quad $ & $%
d_{f}\left( \ell =50\right) \quad $ & $d_{f}\left( \ell =1\right) $ (shuf.)
\quad & $d_{f}\left( \ell =1\right) $ (rand)\quad \\ \hline
aa & 1.45 & 1.42 & 1.43 & 1.45 & 1.45 & 1.68 \\ 
aig & 1.28 & 1.25 & 1.26 & 1.29 & 1.40 & 1.67 \\ 
axp & 1.46 & 1.44 & 1.45 & 1.44 & 1.45 & 1.67 \\ 
ba & 1.44 & 1.45 & 1.43 & 1.44 & 1.45 & 1.67 \\ 
c & 1.33 & 1.31 & 1.31 & 1.31 & 1.33 & 1.67 \\ 
cat & 1.37 & 1.37 & 1.36 & 1.41 & 1.41 & 1.67 \\ 
dd & 1.45 & 1.47 & 1.47 & 1.48 & 1.47 & 1.66 \\ 
dis & 1.50 & 1.50 & 1.50 & 1.51 & 1.51 & 1.66 \\ 
ge & 1.52 & 1.53 & 1.54 & 1.51 & 1.52 & 1.66 \\ 
gm & 1.46 & 1.43 & 1.46 & 1.45 & 1.46 & 1.67 \\ 
hd & 1.37 & 1.39 & 1.37 & 1.39 & 1.36 & 1.68 \\ 
hon & 1.48 & 1.46 & 1.48 & 1.47 & 1.48 & 1.67 \\ 
hpq & 1.46 & 1.46 & 1.47 & 1.47 & 1.50 & 1.67 \\ 
ibm & 1.43 & 1.47 & 1.44 & 1.45 & 1.45 & 1.64 \\ 
intc & 1.35 & 1.40 & 1.40 & 1.43 & 1.41 & 1.69 \\ 
jnj & 1.36 & 1.36 & 1.37 & 1.37 & 1.41 & 1.65 \\ 
jpm & 1.44 & 1.44 & 1.41 & 1.44 & 1.43 & 1.67 \\ 
ko & 1.46 & 1.44 & 1.45 & 1.44 & 1.45 & 1.67 \\ 
mcd & 1.44 & 1.45 & 1.43 & 1.44 & 1.45 & 1.67 \\ 
mmm & 1.33 & 1.31 & 1.31 & 1.31 & 1.33 & 1.67 \\ 
mo & 1.37 & 1.37 & 1.36 & 1.41 & 1.41 & 1.67 \\ 
mrk & 1.45 & 1.47 & 1.47 & 1.48 & 1.47 & 1.66 \\ 
msft & 1.50 & 1.50 & 1.50 & 1.51 & 1.51 & 1.66 \\ 
pfe & 1.52 & 1.53 & 1.54 & 1.51 & 1.52 & 1.66 \\ 
pgn & 1.46 & 1.43 & 1.46 & 1.45 & 1.46 & 1.67 \\ 
sbc & 1.37 & 1.39 & 1.37 & 1.39 & 1.36 & 1.68 \\ 
utx & 1.48 & 1.46 & 1.48 & 1.47 & 1.48 & 1.67 \\ 
vz & 1.46 & 1.46 & 1.47 & 1.47 & 1.50 & 1.67 \\ 
wmt & 1.43 & 1.47 & 1.44 & 1.45 & 1.45 & 1.64 \\ 
xom & 1.35 & 1.40 & 1.40 & 1.43 & 1.41 & 1.69 \\ \hline
\end{tabular}%
\end{table}

\section{Final remarks\label{remarks}}

To summarise, in this manuscript we have made an exhaustive analysis of the
effective multifractal properties of high-frequency price fluctuations and
instantaneous volatility of the equities that compose Dow Jones Industrial
Average. This analysis has comprised the quantification of dependence and
non-Gaussianity on the multifractal character of price fluctuations and
volatility. Furthermore, we have studied the multifractal properties of the $%
\ell $-diagrams made from price fluctuations time series. Our results
indicate that dependence and non-Gaussianity have similar weights on the
multifractal features of both financial quantities. Contrarily to some
stylised facts, and especially for instantaneous volatility \cite%
{cps-volatility}, we have not verified a solid asymptotic power-law decay of
the probability density functions, \textit{i.e.}, fair deviations from
exponential decay. This result is substantiated by the clear approach of $%
\tau \left( z\right) $ curves to the theoretical curve of a independent
gaussian signal when we perform a shuffling on time series elements. If we
consider persistence as a major factor for multiscaling, it might be
puzzling to verify that multifractality for price fluctuations is stronger
than it is for magnitude price fluctuations. Such an apparent contradiction
is cleared up if we take into consideration that price fluctuations PDF
appears to be more fat tailed than instantaneous volatility which introduces
a larger contribution to multiscaling. Besides, in respect of probability
density functions, we have observed that a superstatistical approach to
price fluctuations appears to be valid as a first approach. Still on
multiscaling, we have tried to appraise the robustness of instantaneous
volatility by means of measuring the effect of its possible multifractal
nature on price fluctuations multifractal properties. Our results have
indicated that the non-Gaussinity of instantaneous volatility (price
fluctuation magnitudes) is the chief element of multifractal properties of
price fluctuations. This occurs because the uncorrelated character of the
signal annihilates the influence of dependences of instantaneous volatility
leading to the non-Gaussianity of latter quantity the chief role of
introducing multifractality on price fluctuations time series. In this
perspective heteroskedastic (\textit{i.e.}, $ARCH$) approaches, within
superstatistics is enclosed, to price fluctuations are validated.

Analysing $\ell $-diagrams obtained from price fluctuations time series we
have got sequences of immediate price fluctuations around Cartesian axes
that are forbidden. We have attributed this fact to transaction costs. We
have also observed that despite the number of negative price fluctuations is
greater than the number of positive price fluctuations, the sum all returns
is in fact positive, which is in accordance with both price fluctuations
skewness and economical evolution. By means of a box counting algorithm we
have computed the fractal dimension of such diagrams. We have verified that
the fractal dimension varies slightly when time ordering is destroyed, and
it is deeply affected by randomisation procedures. This provides an
important clue on the fundamental role of non-Gaussinity of price
fluctuations in several properties usually observed.

\subsection*{Acknowledgements}

We would like to thank \textsc{L.G. Moyano} who has performed the intraday
pattern removal described in Sec. \ref{data-method}, as well as \textsc{%
E.M.F. Curado} and \textsc{%
C. Tsallis} for their several comments on the matters which are enclosed in
this manuscript.  One of us (JdS) acknowledges support of \textsc{S.P.
Rostirolla} and \textsc{F. Mancini}. The code for estimating multifractal
spectrum of time series was written during JdS visit to The Abdus Salam
International Centre for Theoretical Physics, Trieste - Italy.
The data used was provided by Olsen Data Services to whom
we are also grateful. We appreciate the useful remarks from \textsc{M.L. Lyra%
} and \textsc{R.L. Viana} at the final stage of the work. This work has
benefited from infrastructural support from PRONEX and PETROBRAS, and
financial support from CNPq (Brazilian agency) and FCT/MCES (Portuguese
agency).

\end{document}